\documentclass[11pt]{article}

\usepackage{a4wide}
\usepackage{amsmath}
\usepackage{bm}
\usepackage{amssymb}
\usepackage{epsfig}
\usepackage{epic}
\usepackage{mathrsfs}
\usepackage[small,bf]{caption}

\newcommand{\insertfig}[2]{\mbox{\epsfxsize=#1cm \epsfbox{#2.eps}}}

\newcommand{\BesselJ}{{J}}

\newcommand{\Tr}{{\rm tr \,}}
\newcommand{\Op}{\mathcal{O}}
\newcommand{\fldZ}{\mathcal{Z}}

\newcommand{\fldD}{\mathcal{D}}

\newcommand{\alg}[1]{\mathfrak{#1}}
\newcommand{\superN}{\mathcal{N}}

\newcommand{\indups}[1]{_{\mathrm{\scriptscriptstyle #1}}}
\newcommand{\sfrac}[2]{{\textstyle\frac{#1}{#2}}}
\newcommand{\EulerGamma}{\gamma_{\indups{E}}}
\newcommand{\pint}{\makebox[0pt][l]{\hspace{2.4pt}$-$}\int}
\newcommand{\rhok}{\rho_{\textrm{\tiny K}}}
\newcommand{\brhok}{\bar{\rho}_{\textrm{\tiny K}}}
\newcommand{\beq}{\begin{equation}}
\newcommand{\eeq}{\end{equation}}

\newcommand{\beqa}{\begin{eqnarray}}
\newcommand{\eeqa}{\end{eqnarray}}

\newcommand{\nn}{\nonumber}

\makeatletter
\def\mr@ignsp#1 {\ifx\:#1\@empty\else #1\expandafter\mr@ignsp\fi}%
\newcommand{\multiref}[1]{\begingroup
\xdef\mr@no@sparg{\expandafter\mr@ignsp#1 \: }%
\def\mr@comma{}%
\@for\mr@refs:=\mr@no@sparg\do{\mr@comma\def\mr@comma{,}\ref{\mr@refs}}%
\endgroup}
\makeatother
\long\def\symbolfootnote[#1]#2{\begingroup%
\def\thefootnote{\fnsymbol{footnote}}\footnote[#1]{#2}\endgroup}

\newcommand{\Appref}[1]{Appendix~\multiref{#1}}

\newcommand{\figref}[1]{figure~\multiref{#1}}
\numberwithin{equation}{section}
\begin{document}
\thispagestyle{empty}

\begin{flushright}\footnotesize
\texttt{AEI-2009-111}\\
\texttt{Imperial-TP-AR-2009-3}\\
\texttt{UUITP-24/09}
\vspace{0.5cm}
\end{flushright}
\setcounter{footnote}{0}

\vspace{6mm}

\begin{center}
{\Large\textbf{\mathversion{bold} From weak coupling to spinning strings}}
\vspace{15mm}

{\sc Lisa Freyhult$^{a,}$\footnote{lisa.freyhult@physics.uu.se}, 
Adam Rej$^{b,}$\footnote{a.rej@imperial.ac.uk} and
Stefan Zieme$^{c,}$\footnote{stzieme@aei.mpg.de}}\\[10mm]

{\it $^a$  Department of Physics and Astronomy, Uppsala University\\
	    P.O. Box 803, S-75108, Uppsala, Sweden}\\[8mm]

{\it $^b$ Blackett Laboratory, Imperial College, London SW7 2AZ, U.K.}\\[8mm]

{\it $^c$ Max-Planck-Institut f\"ur Gravitationsphysik\\
    Albert-Einstein-Institut \\
    Am M\"uhlenberg 1, D-14476 Potsdam, Germany}\\[28mm]

\textbf{Abstract}\\[6mm]
\end{center}

\noindent{We identify the gauge theory dual of a spinning string of minimal energy with 
spins $S_1, S_2$ on $AdS_5$ and charge $J$ on $S^5$. For this purpose we focus on a 
certain set of local operators with two different types of covariant derivatives acting 
on complex scalar fields. We analyse the corresponding nested
Bethe equations for the ground states in the limit of large spins. The auxiliary Bethe 
roots form certain string configurations in the complex plane, which enable us to 
derive integral equations for the leading and sub-leading contribution to the 
anomalous dimension. The results can be expressed through the observables of the 
$\mathfrak{sl}(2)$ sub-sector, i.e.~the cusp anomaly $f(g)$ and the virtual scaling 
function $B_L(g)$, rendering the strong-coupling analysis straightforward.
Furthermore, we also study a particular sub-class of these operators specialising to a 
scaling limit with finite values of the second spin at weak and strong coupling.}
\newpage
\setcounter{footnote}{0}
\setcounter{page}{1}
\section{Introduction and Summary}

Twist operators have so far played a major role in dynamical tests of the AdS/CFT 
correspondence in the planar limit. The main reason is their special scaling 
property at large values 
of the Lorentz spin. In particular, the anomalous dimension grows logarithmically 
with the spin, \textit{cf.}~\cite{BelGorKor06}. This scaling behaviour is not 
a unique feature of the maximally supersymmetric Yang-Mills theory in four dimensions, 
but rather a special case of the so-called Sudakov scaling \cite{Collins:1989bt} 
common to many gauge theories. 

A typical representative of these operators 
in the $\mathfrak{sl}(2)$ sector of $\superN=4$ SYM is built from $L$ complex scalar fields, 
$\fldZ$, and $M$ covariant light-cone derivatives, $\fldD$, acting on the scalar 
background fields within the trace 
\begin{equation}\label{twist op}
\Tr \fldD^M \fldZ^L+\ldots \,.
\end{equation}
The anomalous dimension of these operators occupy a band \cite{BelKorPas09}.
A distinguished sub-set among these is formed by the ground states, 
i.e. the ~lowest operators in the band, which enjoy several additional symmetry properties. 
For the two lowest possible values of the length, $L=2$ and for the ground state of $L=3$\,, analytic 
expressions for the anomalous dimensions can be found 
at high orders in perturbation theory \cite{KotRejZie09, BecBelKotZie09}. 

For twist-two operators 
these coincide with the maximal transcendental terms \cite{KotLip03}
of the known QCD results up to three-loop order, see \cite{MocVerVog04} and 
references therein. At four-loop order the splitting functions of QCD are unknown, 
nevertheless the anomalous dimension of these operators may be determined  
\cite{KotLipRejStaVel07, BajJanLuk08}. The result agrees with constraints  
from the BFKL equation \cite{KotLip03}. 
Moreover, the result for $M=2$ coincides with the explicit 
Feynman diagram computations of \cite{FiaSanSieZan07}. 

The ground 
states for $L=3$ enjoy a similar solvability and the leading wrapping correction may 
be explicitly found \cite{BecForLukZie09}. In the special case of two excitations, 
the result is confirmed by the super-graph computation of \cite{FiaSanSie09}. 

For $L>3$ it is unknown whether closed expressions for the anomalous dimensions can be found,
 even for the ground states. Nevertheless, the anomalous dimensions of the latter 
enjoy very interesting scaling properties in the limit $M \to \infty$,
\beq \label{scaling}
 \gamma_{\mathrm{\scriptscriptstyle sl(2)}}(L,M)=f(g) (\log M+\EulerGamma-(L-2)\log 2 )+B_L(g) + 
		      \Op\big(\tfrac{1}{\log{M}}\big) \,.
\eeq
The function $f(g)$ is also referred to as the cusp anomalous dimension. It is 
conjectured to be independent of the length $L$  and consequently not influenced 
by wrapping interactions.
Thus, one can use the asymptotic Bethe equations to derive an integral equation 
\cite{BeiEdeSta07}, which allows to compute $f(g)$ 
to arbitrary loop order. The scaling function resulting from the weak-coupling 
solution of this equation coincides up to four-loop order with the explicit 
perturbative computations of \cite{Bern:2006ew}. At strong coupling, the solution to 
this integral equation \cite{BasKorKot08,KosSerVol08} 
leads to a remarkable agreement with the corresponding string theory results, 
see \cite{Gubser:2002tv, Frolov:2002av, Roiban:2007ju}. Hence $f(g)$ embodies 
the first known interpolating function of AdS/CFT. A special phenomenological 
interest in this object 
is due to its appearance in multi-loop gluon scattering amplitudes as well as in
expectation values of certain Wilson lines. That is, the scaling
function $f(g)$ determines the leading $1/\epsilon^2$ pole structure of the logarithm
of gluon amplitudes \cite{Bern:2006ew} as well as the logarithmic growth of the 
anomalous dimension of light-like Wilson loops with a cusp \cite{Korchemsky:1985xj},
as first noted in the strong coupling limit \cite{Alday:2007hr}.

The virtual scaling function, $B_L(g)$, appearing in \eqref{scaling} 
explicitly depends on the twist $L$ and it is less obvious that it remains 
unaffected by wrapping effects. The integral equation corresponding to $B_L(g)$
has been derived in \cite{FreZie09}, see also \cite{Freyhult:2007pz, FioGriRos08, 
FioGriRos09}. Interestingly, 
the solution to this equation may be related to the solution of the integral equation 
determining $f(g)$. This intertwines the strong coupling analysis of both 
functions and the methods developed for the cusp anomalous dimension may be directly 
applied also to the case of the virtual scaling function. The resulting strong-coupling 
expansion \cite{FreZie09} is in perfect agreement with the string theory predictions at 
leading and next-to-leading order in $\lambda$, see \cite{BecForTirTse08}, suggesting 
that the wrapping interactions can be neglected also for the first finite-spin 
corrections. Also this quantity appears in the context of gluon amplitudes
and Wilson loops. It enters the sub-leading $1/\epsilon$ poles as part of the collinear
anomalous dimension, see \cite{DixMagSte08}. 

In this paper we will go beyond the $\mathfrak{sl}(2)$ sector. We introduce and investigate 
a gauge theory dual of a minimal energy spinning 
string configuration with two spins, $S_1$ and $S_2$, in $AdS_5$ and charge $J$ in $S^5$. 
The field content of these operators can be schematically represented by
\beq\label{mn Operator}
\Tr \fldD^{n+m} \dot{\bar{\fldD}}^{m} \fldZ^{L} \,.
\eeq
The charges of the string are related to $m$ and $n$ through 
the identification
\beq
S_1=n+m-\tfrac{1}{2}\,,\qquad S_2=m-\tfrac{1}{2}\,,\qquad J=L\,.
\eeq
At weak coupling we extensively examine the limit $m, n \to \infty$ with 
$n/m=\alpha$ fixed, in which $S_1, S_2 \to \infty$ with 
$S_1/S_2=1+\alpha$ fixed. We start with the analysis of the corresponding 
nested one-loop Bethe equations. Surprisingly, the states with minimal length, $L=3$, 
are again solvable and the respective Baxter functions may be found. While we 
extensively use this analytic solution for a numerical study of the behaviour of 
the one-loop Bethe roots in the large $m, n$ limit, we defer its derivation to 
Appendix \ref{app:A}. We find that the auxiliary roots may be decoupled at the first 
two orders in $m, n \gg 1$, leaving a remainder in the main equation. Upon 
introducing the density of roots, this effective equation may be turned into a solvable 
integral equation. Subsequently, we use its leading one-loop solution as the starting 
point for the derivation of the all-loop integral equation for the density. 
The solution to this all-loop equation allows to determine the corresponding anomalous 
dimension. For the first two orders in $m,n$, we find
\beq \label{mainresult}
\gamma(L,m,\alpha) = \frac{3}{2} \, \gamma_{\mathrm{\scriptscriptstyle sl(2)}}(\tfrac{2}{3} \, L,m)
 +\frac{f(g)}{2} \log\Big(\frac{1}{2}(1+\alpha)(2+\alpha)\Big)+
 \Op\Big(\frac{1}{\log{m}}\Big) \,.
\eeq
This result is very surprising, since it suggests a deep relation between the one-spin 
and two-spin solutions at the first two leading orders. This unexpected link calls 
for further investigation on the string theory side.

At leading order in spin the result $\sim \sfrac{3}{2}f(g)$ agrees with the energy 
scaling of spiky-strings \cite{Kruczenski:2004wg, Dorey:2008vp} that 
consist of three arcs, each of them contributing a factor of $\sfrac{1}{2}$, see
\cite{Alday:2007mf,Kruczenski:2008bs}. At finite 
order the results differ \cite{BecForTirTse08,FreKruTir09}.

Very recently it was shown by A. Tirziu and A. Tseytlin 
that in the case of $\alpha=0$ the string theory calculation agrees with the formula 
\eqref{mainresult} at leading order in large $\lambda$, see \cite{TirTse09}. This gives 
first independent evidence\footnote{It would be interesting to calculate the first quantum correction.} 
of absence of wrapping effects in our gauge theory calculation and justifies 
the use of the asymptotic Bethe equations.

We furthermore examine the large $n$ limit of \eqref{mn Operator} for arbitrary finite
values of $m \geq2$. The strong coupling solution cannot be expressed in terms
of known observables. However, the same methodology as in \cite{FreZie09} can be used, to 
solve the corresponding integral equation at strong coupling. At leading order in $n$ this 
solution resembles the folded string configuration.

\section{Definitions and the string theory dual}
\label{weak g}
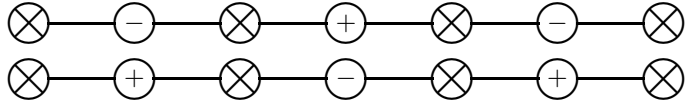
\begin{figure}\centering
\begin{minipage}{260pt}
\setlength{\unitlength}{1pt}%
\small\thicklines%
\begin{picture}(260,20)(-10,-10)
\put(  0,00){\circle{15}}%
\put(  7,00){\line(1,0){26}}%
\put( 40,00){\circle{15}}%
\put( 47,00){\line(1,0){26}}%
\put( 80,00){\circle{15}}%
\put( 87,00){\line(1,0){26}}%
\put(120,00){\circle{15}}%
\put(127,00){\line(1,0){26}}%
\put(160,00){\circle{15}}%
\put(167,00){\line(1,0){26}}%
\put(200,00){\circle{15}}%
\put(207,00){\line(1,0){26}}%
\put(240,00){\circle{15}}%
\put( -5,-5){\line(1, 1){10}}%
\put( -5, 5){\line(1,-1){10}}%
\put( 75,-5){\line(1, 1){10}}%
\put( 75, 5){\line(1,-1){10}}%
\put(155,-5){\line(1, 1){10}}%
\put(155, 5){\line(1,-1){10}}%
\put(235,-5){\line(1, 1){10}}%
\put(235, 5){\line(1,-1){10}}%
\put( 40,00){\makebox(0,0){$-$}}%
\put(120,00){\makebox(0,0){$+$}}%
\put(200,00){\makebox(0,0){$-$}}%
\end{picture}
\end{minipage}
\begin{minipage}{260pt}
\setlength{\unitlength}{1pt}%
\small\thicklines%
\begin{picture}(260,20)(-10,-10)
\put(  0,00){\circle{15}}%
\put(  7,00){\line(1,0){26}}%
\put( 40,00){\circle{15}}%
\put( 47,00){\line(1,0){26}}%
\put( 80,00){\circle{15}}%
\put( 87,00){\line(1,0){26}}%
\put(120,00){\circle{15}}%
\put(127,00){\line(1,0){26}}%
\put(160,00){\circle{15}}%
\put(167,00){\line(1,0){26}}%
\put(200,00){\circle{15}}%
\put(207,00){\line(1,0){26}}%
\put(240,00){\circle{15}}%
\put( -5,-5){\line(1, 1){10}}%
\put( -5, 5){\line(1,-1){10}}%
\put( 75,-5){\line(1, 1){10}}%
\put( 75, 5){\line(1,-1){10}}%
\put(155,-5){\line(1, 1){10}}%
\put(155, 5){\line(1,-1){10}}%
\put(235,-5){\line(1, 1){10}}%
\put(235, 5){\line(1,-1){10}}%
\put( 40,00){\makebox(0,0){$+$}}%
\put(120,00){\makebox(0,0){$-$}}%
\put(200,00){\makebox(0,0){$+$}}%
\end{picture}
\end{minipage}
\caption{Dynkin diagrams of $\alg{su}(2,2|4)$ with different gradings \cite{Beisert:2005fw}.}\label{Dynkindiagram}
\end{figure}

The highest-weight states corresponding to  operators \eqref{mn Operator} have the 
following Dynkin labels with respect to the upper Dynkin diagram in \figref{Dynkindiagram}, \textit{cf.} 
\cite{Beisert:2005fw},
\beq \label{dynlab}
(\Delta_0, s_1, s_2, q_1, p, q_2, B, L) 
=(L+2m+n-\tfrac{3}{2}, 2m+n-1, n, 1, L-3, 2, \tfrac{1}{2}, L)\,.
\eeq

This suggests that the string theory dual of these operators is a spinning string 
with two spins $S_1, S_2$ on $AdS_5$ and charge $L$ on the $S^5$. Upon the 
usual $SO(4)$ rotation, we can read of the spins directly from \eqref{dynlab} 
\beq \label{stringspins}
S_1=m+n-\tfrac{1}{2}\,, \qquad S_2=m-\tfrac{1}{2}\,.
\eeq
In view of the comparison with the string results, we will consider 
large values of $m$ and $n$ while fixing their ratio to $\frac{n}{m}=\alpha$. 
For finite values of $\alpha$ both spins become large and
\beq 
\frac{S_1}{S_2}=1+\alpha\,.
\eeq
In particular, the case of $\alpha=0$ corresponds to the symmetric case of the spinning 
string with equal spins. The well-studied case of the folded string with one large spin,
on the other hand, corresponds to $\alpha=\infty$. In this publication, we will only 
study the gauge theory states with minimal anomalous dimension, which correspond to 
minimal energy states on the string theory side.

In contradistinction to the $\mathfrak{sl}(2)$ sub-sector, the Bethe equations 
corresponding to the operators \eqref{mn Operator} are nested. The number of nesting
levels depends on the choice of representation. The minimal 
number of levels is equal to three. However, for the sake of convenience,  
we will mostly use a non-minimal representation corresponding to the lower Dynkin 
diagram in \figref{Dynkindiagram}.

A subset of \eqref{mn Operator} with $L=3$ has already been 
studied in the literature, \textit{cf.}~\cite{BeiBiaMorSam04}. It was found 
experimentally that the corresponding one-loop anomalous dimension is given by the 
following closed formula
\begin{equation}\label{eq:one loop gamma}
\gamma_{n,m} = 4\,H_1\left(\tfrac{m}{2}-\tfrac{1}{2}\right)
+4\,H_1\left(m+\tfrac{n}{2}\right)
+4\,H_1\left(\tfrac{m}{2}+\tfrac{n}{2}\right)
-4\,H_1\left(-\tfrac{1}{2}\right) \, .
\end{equation}
Here, $H_1 (N)$ denotes $N$-th harmonic number. We will prove this formula in 
Appendix \ref{app:A}, by constructing an explicit one-loop solution for $L=3$. 
This is a counterpart of the $L=2$ solution found in the $\mathfrak{sl}(2)$ sector 
\cite{Eden:2006rx}. The corresponding Baxter functions (see \eqref{Q4sol}, \eqref{Q5sol}
and \eqref{Q6sol}) become quite complicated and are not hypergeometric orthogonal 
polynomials anymore. Nevertheless, the explicit one-loop solution enormously 
facilitates numerical studies of the Bethe solutions at large values of $m$ and $n$.

\section{Large spins solution at one-loop order}
In this section we will analyse the operators \eqref{mn Operator} in the limit 
$m,n \to \infty$ and $n/m= \alpha=\textrm{fixed}$ at leading order in perturbation 
theory. As in the case of the $\mathfrak{sl}(2)$ operators \cite{Eden:2006rx}, the 
leading solution \textit{does not} depend on the length $L$. The minimal set of 
equations at one-loop consists of two coupled nesting levels
\beqa
\left(\frac{u_{4,k}+\frac{i}{2}}{u_{4,k}-\frac{i}{2}}\right)^L&=&
\prod^{n+2m}_{\substack{j=1 \\ j\neq k}} \frac{u_{4,k}-u_{4,j}-i}{u_{4,k}-u_{4,j}+i}
\prod^{m}_{j=1} \frac{u_{4,k}-u_{5,j}+\frac{i}{2}}{u_{4,k}-u_{5,j}-\frac{i}{2}} 
\label{main1}\\
1&=&\prod^{m}_{\substack{j=1 \\ j\neq k}} \frac{u_{5,k}-u_{5,j}-i}{u_{5,k}-u_{5,j}+i} 
\prod^{n+2m}_{j=1} \frac{u_{5,k}-u_{4,j}+\frac{i}{2}}{u_{5,k}-u_{4,j}-\frac{i}{2}} 
\label{main2}\,.
\eeqa
From the numerical studies of the analytic solution for $L=3$ presented in 
\Appref{app:A} we infer that the roots $u_{5}$ form strings
\begin{eqnarray}\label{u5roots}
\nonumber u_5&\to& \pm i/2,\pm 3i/2, \pm \dots \quad\mbox{$m$ even},\\
u_5&\to& 0,\pm i,\pm2i,\pm \dots \quad\mbox{$m$ odd}.
\end{eqnarray}
However, it is \textit{incorrect} to assume, even at the leading order, that all 
$u_5$ will exhibit this behaviour. We thus introduce an effective 
cut-off $c(\alpha)$ such that
\begin{eqnarray}\label{effstack}
\nonumber u_5&\to& \pm i/2,\pm 3i/2, \dots, \pm i (m/c(\alpha)-1/2)
\quad\mbox{$m$ even},\\
u_5&\to& 0,\pm i,\pm2i, \dots,  \pm i (m/c(\alpha)-1/2) \quad  \mbox{$m$ odd}.
\end{eqnarray}
In the following we will explicitly fix $c(\alpha)$ in the limits $m, n \to \infty$.
We have also checked numerically that the remaining roots, i.e.~not belonging 
to the effective strings \eqref{effstack}, scale as $\sim m^2$ at large values of $m$ 
and thus are not relevant for the leading and the sub-leading order. In the large $n$ 
limit and for finite values of $m$ the strings \eqref{u5roots}, as we will show in what 
follows, become exact and $c(\infty)=2$ as expected. In either case, one can 
completely decouple the $u_5$ roots. The effective equation for the 
middle-node roots thus takes the following form
\beq
\left(\frac{u_{4,k}+\frac{i}{2}}{u_{4,k}-\frac{i}{2}}\right)^L=
\left(\frac{u_{4,k}+\frac{i \, m}{c(\alpha)}}{u_{4,k}-\frac{i\, m}{c(\alpha)}} \right)
\prod^{m(\alpha+2)}_{\substack{j=1 \\ j\neq k}} 
\frac{u_{4,k}-u_{4,j}-i}{u_{4,k}-u_{4,j}+i}\,.
\label{main1decouple}\\
\eeq
In the large $m$ and/or $n$ limit this equation may be turned into an integral equation along 
the lines presented in \cite{Eden:2006rx}. Explicitly, one obtains
\beqa  \nn
\frac{2\,L}{m(\alpha+2)}\arctan(2u_k)&=&\frac{2\,\pi}{m(\alpha+2)} \widetilde{n}(u)
-2\int_{-b(\alpha)}^{b(\alpha)} du' \rho(u') \arctan(u-u')\\
\label{inteq}
&&+\frac{2}{m(\alpha+2)}\arctan(\tfrac{c(\alpha) \,u_k}{m})\,.
\eeqa
Here, $\widetilde{n}(u)$ is the function for the mode numbers. It is related to the 
density $\rho_0(u)$ through
\beq
\widetilde{n}(u)=\frac{L-3}{2}\textrm{sgn}(u)-\frac{m(\alpha+2)}{2}
+\int^{u}_{-b(\alpha)}du' \rho(u')\,.
\eeq
%
\subsection{The $\alpha=0$ solution}
Upon rescaling the momentum-carrying roots $u_4 = 2m\, \bar{u}_4$ and the 
corresponding density $\bar{\rho}_0(\bar{u})=2m \,\rho_0 (u)$, the leading part of 
the equation \eqref{inteq} is given by
\begin{equation}
3\pi \,\textrm{sgn}(\bar{u}) - 2\,\arctan(2\,c\, \bar{u})
+2\pint_{-\bar{b}(0)}^{\bar{b}(0)}d\bar{u}' \frac{\bar{\rho}_0(\bar{u}')}{(\bar{u}-\bar{u}')}=0\,.
\end{equation}
The solution to this equation is given by
\begin{equation}\label{largemoneloop}
\bar{\rho}_0(\bar{u})=\frac{3}{2} \, \brhok \Big(\frac{\bar{u}}{2\, \bar{b}(0)}\Big)-
\frac{1}{2\pi}\log \left(\frac{\sqrt{1-\frac{\bar{u}^2}{\bar{b}(0)^2}}+
\frac{\sqrt{4\bar{b}(0)^2 \,c^2+1}}{2 \bar{b}(0) c}}
{\frac{\sqrt{4\bar{b}(0)^2 c^2+1}}{2 \bar{b}(0) c}
-\sqrt{1-\frac{\bar{u}^2}{\bar{b}(0)^2}}}\right)\,.
\end{equation}
Here, with $\rhok(u)$ we have denoted the one-loop density of the Bethe roots 
corresponding to the ground state of twist-$L$ operators in $\mathfrak{sl}(2)$. 
The normalisation condition
\beq
1=\int^{\bar{b}(0)}_{-\bar{b}(0)}d\bar{u}\, \bar{\rho}_0(\bar{u})\,,
\eeq
relates the boundary of the root distribution to the dimension of the effective 
strings $c$. Explicitly, one finds
\beq \label{b0}
\bar{b}(0) = \frac{1}{16\,c}( \sqrt{9 - 4 \,c + 4\, c^2} + 6 \,c-3)\,.
\eeq
To determine the constant $c$ we proceed as follows. The density $\rho_0(u)$ has the 
following large $m$ expansion
\beq \label{rho0exp}
\rho_0(u)=\frac{3}{4m}\left(\frac{2}{\pi}\log m+C-\frac{2}{\pi^2}\log(u^2)\right)+\dots\,,
\eeq
with the constant term $C$ given by
\beq \label{calphazero}
C=\frac{2}{\pi} \log \left(4 \, \bar{b}(0) \right)-\frac{2}{3 \pi }\log{\left(2\, \bar{b}(0)\,  c+\sqrt{4\,\bar{b}(0)^2 \,c^2+1}\right)}\,.
\eeq
One may thus split the density in \eqref{inteq} as follows
\beq \label{rhosplit}
\rho(u) = \rho_0(u) + r(u)\, ,
\eeq
and use the expansion \eqref{rho0exp} to obtain a leading integral equation for 
$r(u)$. Upon Fourier transformation one finds
\beq \label{ra0}
\hat{r}(t)=\frac{1}{2m}\left(3\frac{e^{-|t|/2}}{1-e^{-|t|}}-\frac{3}{|t|}-\frac{L}{1+e^{-|t|/2}}\right)\,.
\eeq
One can now use \eqref{rhosplit} to calculate the one-loop anomalous dimension. 
After straightforward integration one finds
\beq
\gamma_0=12\log m+12\EulerGamma-8(L-3)\log 2+6\,\pi\,C.
\eeq
This should be compared with the large $m$ expansion of the $L=3$ analytic result 
(\ref{eq:one loop gamma})
\beq
\gamma^{L=3}_0=12\log m+12\EulerGamma \,.
\eeq
Thus, the constant $c$ is determined by the condition $C=0$ in conjunction with 
\eqref{b0}. Numerically one may determine $c=2.83181(\dots)$, which means that 
approximately $\sfrac{7}{10}$ of the $u_5$ roots form effective strings. We have checked 
numerically that $\bar{b}(0)$ obtained by inserting this value of $c$ is consistent with 
the scaling properties of the largest $u_4$ roots.
%
\subsection{The solution for general $\alpha>0$}
\begin{figure}
\begin{center}

\mbox{
\begin{picture}(0,150)(220,0)
\put(130,0){\insertfig{5.5}{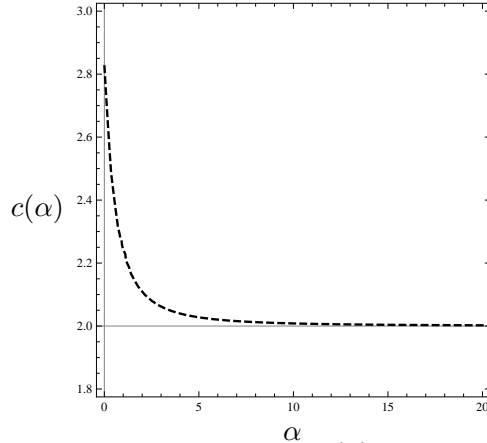}}
\put(105,75){$c(\alpha)$}
\put(208,-10){$\alpha$}
\end{picture}}
\caption{The plot of $c(\alpha)$ as function of $\alpha$.} \label{calphaplot}
\end{center}
\end{figure}

The limit $m \to \infty, n \to \infty$ with $\frac{n}{m}=\alpha=\textrm{fixed}$ is a 
simple generalisation of the $\alpha=0$ case discussed above. The roots 
$u_5$ form again strings, and one expects the cut-off parameter to depend 
on $\alpha$, i.e. $c=c(\alpha)$. Upon rescaling the roots by $n+2m= m(\alpha+2)$, 
one derives the following integral equation
\begin{equation}
3\pi \,\textrm{sgn}(\bar{u}) - 2\,\arctan((2+\alpha)\,c(\alpha)\, \bar{u})+2\pint_{-\bar{b}(\alpha)}^{\bar{b}(\alpha)}d\bar{u}' \frac{\bar{\rho}_0(\bar{u}',\alpha)}{(\bar{u}-\bar{u}')}=0\,.
\end{equation}
The solution to this equation is given by
\begin{equation}\label{alphaoneloop}
\bar{\rho}_0(\bar{u},\alpha)=
\frac{3}{2} \, \brhok \Big(\frac{\bar{u}}{2\, \bar{b}(\alpha)} \Big)
-\frac{1}{2 \pi}\log \left(\frac{\sqrt{1-\frac{\bar{u}^2}{\bar{b}(\alpha)^2}}+\frac{\sqrt{\bar{b}(\alpha)^2 (\alpha +2)^2 c(\alpha
   )^2+1}}{\bar{b}(\alpha) (\alpha +2) c(\alpha )}}{\frac{\sqrt{\bar{b}(\alpha)^2 (\alpha +2)^2 c(\alpha )^2+1}}{\bar{b}(\alpha)
   (\alpha +2) c(\alpha )}-\sqrt{1-\frac{\bar{u}^2}{\bar{b}(\alpha)^2}}}\right)\,.
\end{equation}
The normalisation condition yields the relation
\begin{equation}\label{normalization}
\bar{b}(\alpha)=\frac{1}{8\,(2+\alpha)\, c(\alpha)}\left(-3+3\,(2+\alpha)\,c(\alpha)+\sqrt{9-2\,(2+\alpha)\,c(\alpha)+(2+\alpha)^2\,c(\alpha)^2}\right).
\end{equation}
The constant $c(\alpha)$ may be determined by the same procedure as in the 
$\alpha=0$ case. The denisty $\rho_0(u)$ has the following large $m$ expansion
\beq
\nonumber\rho_0(u,\alpha)=\frac{3}{2m(2+\alpha)}\left(\frac{2}{\pi}\log m+C(\alpha)-\frac{2}{\pi^2}\log(u^2)\right)\,,
\eeq
with
\beq \label{calpha}
C(\alpha)=\frac{2}{\pi} \log \left(2 \, \bar{b}(\alpha) (\alpha +2) \right)
-\frac{2}{3 \pi }\log{\left(\bar{b}(\alpha) (\alpha +2) c(\alpha )+\sqrt{\bar{b}(\alpha)^2 (\alpha +2)^2 c(\alpha
   )^2+1}\right)}\,.
\eeq
Splitting again the leading density as
\beq
\rho(u,\alpha)=\rho_0(u,\alpha)+r(u,\alpha)+\ldots \,,
\eeq
one determines $\hat{r}(t,\alpha)$ to be
\beq
\hat{r}(t,\alpha)=\frac{1}{m(2+\alpha)}\left(3\frac{e^{-|t|/2}}{1-e^{-|t|}}-\frac{3}{|t|}-\frac{L}{1+e^{-|t|/2}}\right)\,.
\eeq
This immediately leads to
\beq
\gamma_0=12\log m+12\EulerGamma-8(L-3)\log 2+6\,\pi\,C (\alpha).
\eeq
Expanding (\ref{eq:one loop gamma}) for general $\alpha$, one finds
\begin{equation} \label{energywithalpha}
\gamma_0=12\log m+12\EulerGamma+4\log\left(\tfrac{1}{2}(1+\alpha)(2+\alpha)\right) \,.
\end{equation}
One thus concludes that
\beq
3\,\pi\,C(\alpha)=2\log\left(\tfrac{1}{2}(1+\alpha)(2+\alpha)\right) \,.
\eeq
The above formula in conjunction with \eqref{calpha} and \eqref{normalization} 
determines $c(\alpha)$ uniquely. The behaviour of $c(\alpha)$ as a function of $\alpha$ 
is shown in figure \ref{calphaplot}. Clearly, the effective strings become exact in 
the large $\alpha$ limit
\beq \label{calpha2}
\lim_{\alpha \to \infty} c(\alpha) \to 2 \,.
\eeq
\subsection{The $\alpha = \infty$ solution} \label{sec:infinitealpha}
In view of \eqref{calpha2}, one concludes that for $\alpha \to \infty$ all roots $u_5$ 
form strings as in \eqref{u5roots}. Rescaling the main roots as $u_4 = n \,\bar{u}_4$ and taking
the large $n$ limit we find
\begin{equation}
0=2\pi\,\textrm{sgn}(\bar{u})
-2\,\pint_{-b/n}^{b/n} d\bar{u}'\frac{\bar{\rho}_0(\bar{u})}{(\bar{u}-\bar{u}')}.
\end{equation}
The above equation, together with the momentum constraint which fixes $b$ to $n/2$, 
is solved by the known $\mathfrak{sl}(2)$ density, $\brhok(\bar{u})$, found in 
\cite{Korchemsky:1995be}. This formally corresponds to taking $\alpha \to \infty$, although
the latter is not unique due to the fact that any dependence on finite values of
the spin $S_2$ is lost. 
\section{The all-loop equations}
Beyond the one-loop order the spectral equations for the operators \eqref{mn Operator} 
can be derived from the full system of the asymptotic Bethe equations conjectured 
in \cite{Beisert:2005fw}. In what follows, we will work with a set of four coupled 
Bethe equations
\begin{eqnarray}\label{sl2BE}
\nonumber\left(\frac{x_{4,k}^+}{x_{4,k}^-}\right)^L
&=&\prod_{j\neq k}^{n+2m}\frac{x_{4,k}^--x_{4,j}^+}{x_{4,k}^+-x_{4,j}^-}
\frac{1-g^2/x_{4,k}^+x_{4,j}^-}{1-g^2/x_{4,k}^-x_{4,j}^+} \, \sigma^2(u_{4,k},u_{4,j})\\
\nonumber&\times&\prod_{j=1}^{m}\frac{x_{4,k}^{+}-x_{5,j}}{x_{4,k}^{-}-{x}_{5,j}}
\prod_{j=1}^{m-2}\frac{1-g^2/x_{4,k}^+{x}_{7,j}}{1-g^2/x_{4,k}^-{x}_{7,j}}\\
\nonumber1&=&\prod_{j=1}^{m-1}\frac{{u}_{5,k}-u_{6,j}-i/2}{{u}_{5,k}-u_{6,j}+i/2}
\prod_{j=1}^{n+2m}\frac{{x}_{5,k}-x_{4,j}^-}{{x}_{5,k}-x_{4,j}^+}\\
\nonumber1&=&\prod_{j=1}^{m-1}\frac{u_{6,k}-u_{6,j}+i}{u_{6,k}-u_{6,j}-i}
\prod_{j=1}^{m}\frac{u_{6,k}-{u}_{5,j}-i/2}{u_{6,k}-{u}_{5,j}+i/2}
\prod_{j=1}^{m-2}\frac{u_{6,k}-{u}_{7,j}-i/2}{u_{6,k}-{u}_{7,j}+i/2}\\
1&=&\prod_{j=1}^{m-1}\frac{{u}_{7,k}-u_{6,j}-i/2}{{u}_{7,k}-u_{6,j}+i/2}
\prod_{j=1}^{n+2m}
\frac{1-g^2/{x}_{7,k}x_{4,j}^-}{1-g^2/{x}_{7,k}x_{4,j}^+}.
\end{eqnarray}
Our convention for the coupling constant is 
$g^2 = g^2_{\rm\scriptscriptstyle YM}N_c/(16 \pi^2)$. The deformation of the spectral
parameter reads $x(u) =\sfrac{1}{2} (u + \sqrt{u^2 -4 g^2})$, with the  
conventional notation $x^\pm = x(u \pm \sfrac{i}{2})$. The form of the dressing
factor $\sigma^2$ is given in \cite{BeiEdeSta07}. Note that at one-loop order, 
the generating functions of the $u_6$ and $u_7$ roots can be immediately obtained from 
the ${u}_5$ roots. Consequently, these roots inherit the behaviour of the 
${u}_5$ roots and also form effective strings. We will denote the effective 
cut-off parameter for the ${u}_7$ roots by $d(\alpha)$.

In the large $m,n$ limit, the leading positions of the inner ${u}_5$ 
roots are again given by \eqref{effstack}. The quantum corrections to these roots 
vanish as $m,n \to \infty$. The outer roots, on the other hand, grow very rapidly.  
The same is true for the auxiliary roots ${u}_7$. Thus, effectively, the 
system \eqref{sl2BE} reduces to 
\begin{eqnarray}\label{redBA}
\left(\frac{x_k^+}{x_k^-}\right)^L &=& \prod_{j\neq k}^{n+2m}
\frac{u_k-u_j-i}{u_k-u_j+i}\left(\frac{1-g^2/x_k^+x_j^-}{1-g^2/x_k^-x_j^+}\right)^2
\nonumber\\
&\times&
\frac{u_k+i\tfrac{m}{c(\alpha)}}{u_k-i\tfrac{m}{c(\alpha)}} \,
\frac{1-g^2/x_k^-x(i\tfrac{m}{c(\alpha)})}{1+g^2/x_k^+x(i\tfrac{m}{c(\alpha)})}\,
\frac{1+g^2/x_k^-x(i\tfrac{m}{d(\alpha)})}{1-g^2/x_k^+x(i\tfrac{m}{d(\alpha)})}
\, \sigma^2(u_k,u_j).
\end{eqnarray}
From this set of effective Bethe equations we will derive an integral for the 
fluctuation density, $\sigma(g,u,\alpha)$,
\beq
\rho(g,u,\alpha) = \rho(u,\alpha)-\frac{8\,g^2}{m(\alpha+2)}\,\sigma(g,u,\alpha)\,.
\eeq
In what follows, we will treat the cases $\alpha=0$, $\alpha>0$ and $\alpha=\infty$ 
separately, even though the first two directly interpolate between each other. The 
case of $\alpha=\infty$ requires a more careful analysis.
\subsection{The equal spin case $\alpha=0$}
Knowing \eqref{rho0exp} and \eqref{ra0}, one can easily derive
the integral equation for the Fourier-Laplace transform\footnote{We define the 
Fourier-Laplace transform of the fluctuation density by 
$\hat{\sigma}(t)=e^{-\frac{t}{2}} \int^{\infty}_{-\infty} du\, e^{-itu} \sigma(u)$.} 
of the fluctuation density $\hat{\sigma}(g,t,0)$
\begin{eqnarray}\label{eq:large m dens}
\hat{\sigma}(g,t,0)&=&\frac{t}{e^t-1}\Big[\left(\tfrac{3}{2}\log m
+\tfrac{3}{2}\EulerGamma-(L-3)\log 2\right)K(2gt,0)
-\frac{L}{8g^2t}\left(J_0(2gt)-1\right)\nonumber\\
&&+\frac{1}{2}\int_0^\infty dt'
\Big(\frac{3}{e^{t'}-1}-\frac{L-3}{e^{t'/2}+1}\Big)
\left(K(2gt,2gt')-K(2gt,0)\right)\nonumber\\
&&-4g^2\int_0^\infty dt'K(2gt,2gt')\hat{\sigma}(g,t',0) 
\Big]\,.
\end{eqnarray}
It is written in terms of the usual integral kernel 
$K(t,t')=K_0(t,t')+K_1(t,t')+K_d(t,t')$, with the parity even and odd components 
given respectively by \cite{BeiEdeSta07} 
\begin{eqnarray}\label{eq:kernels}
 K_0(t,t')&=&
      \frac{t \BesselJ_1(t)\BesselJ_0(t') - t'\BesselJ_0(t)\BesselJ_1(t')}{t^2-t'^2}
=\frac{2}{t t'} \sum_{n=1}^{\infty} 
	    (2n-1)\BesselJ_{2n-1}(t) \BesselJ_{2n-1}(t') 
\,,\nonumber\\
 K_1(t,t')&=&
	\frac{t'\BesselJ_1(t)\BesselJ_0(t') - t \BesselJ_0(t)\BesselJ_1(t')}{t^2-t'^2}
=\frac{2}{t t'} \sum_{n=1}^{\infty} (2n)\BesselJ_{2n}(t)\BesselJ_{2n}(t')
\,.
\end{eqnarray}
The dressing kernel $K_d(t,t')$ is a convolution of the even 
and the odd part
\begin{equation}
 K_d(t,t)=8g^2 \int_{0}^{\infty}dt'' K_1(t,2gt'')\frac{t''}{e^{t''}-1}K_0(2gt'',t')\,.
\end{equation}
The anomalous dimension corresponds to the value of $\hat{\sigma}(g,t,0)$ at the origin
\begin{equation} \label{gammathroughsigma}
 \gamma(L,m)=16\,g^2\,\hat{\sigma}(g,0,0) \,.
\end{equation}
By comparing \eqref{eq:large m dens} with the corresponding equation for 
the $\mathfrak{sl}(2)$ sector \cite{FreZie09}, one easily infers that terms 
proportional to $K(2gt,0)$ in the first line of \eqref{eq:large m dens} give rise 
to the cusp anomalous dimension $f(g)$, while the remaining terms yield the 
virtual scaling function $B_2(g)$ obtained in \cite{FreZie09}, 
and the first generalised scaling function $\epsilon_1(g)$ of 
\cite{BasKor09,Fioravanti:2008rv}.
Hence, the anomalous dimension is given by
\beqa \nn
\gamma(L,m)&=&\frac{3}{2}f(g)\left(\log m+\EulerGamma\right)
+(L-3)\epsilon_1(g)+\frac{3}{2}B_2(g)+\Op\Big(\frac{1}{\log{m}}\Big)\ \\ \label{eq:large m gamma}
&=&\frac{3}{2} \, \gamma_{\mathrm{\scriptscriptstyle sl(2)}}(\tfrac{2}{3}L,m)
+\Op\Big(\frac{1}{\log{m}}\Big)\,.
\eeqa
Remarkably, the large $m$ anomalous dimension is up to the order 
$\Op ( \sfrac{1}{\log{m}})$ proportional to the anomalous dimension of the 
twist operators. This can be traced back to the ``screening properties'' of the 
${u}_5$ roots, \textit{cf.}~\eqref{effstack}. It should be noted that 
although we have assumed $\alpha=0$, one does not need to put $n=0$. In contrary, 
\eqref{eq:large m gamma} is valid also in the latter case and the first finite $n$ 
corrections are sub-leading. We expect, in similarity to the case of $\mathfrak{sl}(2)$ 
operators, that the wrapping corrections will not affect the first two orders in the 
large spin, $S=2S_1$, expansion. The findings of \cite{TirTse09} confirm this 
hypothesis.
\subsection{The case of $\alpha>0$}
It is straightforward to repeat the computation of the preceding paragraph for 
$\alpha>0$, the only difference being an additional one-loop term in the energy, 
\textit{cf.}~\eqref{energywithalpha}. The resulting equation for 
$\hat{\sigma}(g,t,\alpha)$ takes the following form
\begin{eqnarray}\label{eq:general alpha dens}
\hat{\sigma}(g,t,\alpha)&=&\frac{t}{e^t-1}\Big[\left(\tfrac{3}{2}\log m
+\tfrac{3}{2}\EulerGamma-(L-3)\log 2+\tfrac{1}{2}\log\left(\tfrac{1}{2}(1+\alpha)(2+\alpha)\right)\right)K(2gt,0)\nonumber\\
&&-\frac{L}{8g^2t}\left(J_0(2gt)-1\right)
+\frac{1}{2}\int_0^\infty dt'
\Big(\frac{3}{e^{t'}-1}-\frac{L-3}{e^{t'/2}+1}\Big)
\left(K(2gt,2gt')-K(2gt,0)\right)\nonumber\\
&&-4g^2\int_0^\infty dt'K(2gt,2gt')\hat{\sigma}(g,t',\alpha) 
\Big]\,.
\end{eqnarray}
The anomalous dimension can be easily found
\beq \label{eq:general alpha gamma}
\gamma_L(g,m)=\frac{3}{2} \gamma_{\mathrm{\scriptscriptstyle sl(2)}}(\tfrac{2}{3}L,m)
+\frac{f(g)}{2}\log\Big(\frac{1}{2}(1+\alpha)(2+\alpha)\Big)
+\Op\Big(\frac{1}{\log{m}}\Big) \,.
\eeq
It is noteworthy that the dependence on $\alpha$ is logarithmic and that the 
corresponding prefactor is again proportional to $f(g)$. It would be interesting 
to understand from a string theory perspective why the energy of the general two-spin 
solution so closely resembles the one-spin solution.
%
\subsection{One large spin limit}
In this section we will discuss the case of $n \to \infty$ and \textit{finite} values 
of $m$. It turns out that, contrary to the $\alpha=0$ case, the sub-leading correction 
exhibits an interesting dependence on $m$.

The one-loop leading solution has been discussed in section \ref{sec:infinitealpha} 
and is fully equivalent to the one-loop problem for the ground states in the 
$\mathfrak{sl}(2)$ sub-sector, \textit{cf.} \cite{Eden:2006rx}. Proceeding in the 
same spirit as in the previous sections, we split off the leading density
\beq
\rho(u) = \rho_0(u)+r(u)
\eeq
and derive a leading equation for $r(u)$, which subsequently may be solved by 
Fourier transformation
\beq\label{eq:LO large n dens}
\hat{r}(t)=
\frac{1}{n}\Big(-\frac{L-3}{1+e^{-|t|/2}}+
3\frac{e^{-|t|/2}}{1-e^{-|t|}}-\frac{e^{-m|t|/2}}{1-e^{-|t|}}-\frac{2}{|t|}\Big).
\eeq
Using this expression, it is straightforward to calculate the one-loop anomalous 
dimension for the first two orders in $n$  
\begin{eqnarray}\label{eq:LO large n E0}
E_0=4\log n+6\EulerGamma+2\psi_0(\tfrac{m+1}{2})-4(L-3)\log 2 \,.
\end{eqnarray}
The derivation of higher-loop corrections goes along similar lines as in the preceding 
sections. Upon defining the fluctuation density by
\beq
\rho(g,u,m) = \rho(u,m)-\frac{8\,g^2}{n}\,\sigma(g,u,m)\,,
\eeq
one derives the following closed integral equation for $m \geq 2$
\begin{eqnarray}\label{eq:large n dense}
\hat{\sigma}(g,t,m)&=&\frac{t}{e^t-1}\Big[\left(\log n+\tfrac{3}{2}\EulerGamma
+\tfrac{1}{2}\psi_0(\tfrac{m+1}{2})-(L-3)\log 2\right)K(2gt,0)
-\frac{L}{8g^2t}\left(J_0(2gt)-1\right)\nonumber\\
&&+\frac{1}{2}\int_0^\infty dt'\Big(\frac{3}{e^{t'}-1}-
\frac{e^{-(m+1)t'/2}}{1-e^{-t'}}-\frac{L-3}{e^{t'/2}+1}\Big)
\left(K(2gt,2gt')-K(2gt,0)\right)\nonumber\\
&&-4g^2\int_0^\infty dt'K(2gt,2gt')\hat{\sigma}(g,t')-
\frac{1}{2}\int_0^\infty K_1(2gt,2gt')e^{-t'\tfrac{m-1}{2}} \,.
\end{eqnarray}
The resulting anomalous dimension scales logarithmically with $n$
\begin{equation}\label{eq:large n gamma}
 \gamma(L,n,m)=16g^2\sigma(g,0,m)=f(g)\log n + \dots \,,
\end{equation}
while the finite-spin corrections depend explicitly on $m$, as can be directly 
inferred from \eqref{eq:large n dense}. At weak-coupling, one can easily determine 
the perturbative expansion at the first few orders 
\begin{eqnarray}\label{gamma large n}
 \gamma(L,n,m)&=&f(g)(\log n + \tfrac{3}{2}\EulerGamma + \tfrac{1}{2}\psi_0(\tfrac{m+1}{2})
-(L-3)\log 2) -2g^4(\psi_2(\tfrac{m+1}{2})+2(21-4L)\zeta(3) )\nonumber\\
&& +\tfrac{1}{3}g^6 \Big( \psi_4(\tfrac{m+1}{2})+2\pi^2 \psi_2(\tfrac{m+1}{2}) 
+24\psi_1(\tfrac{m+1}{2})\zeta(3)+\tfrac{96}{(m-1)^2} \zeta(3) \nonumber\\
&& + 8(6-L)\pi^2\zeta(3)+72(31-7L)\zeta(5)\Big) +\dots\,,
\end{eqnarray}
For the choice of parameters $m=2$ and $L=3$ we find 
perfect agreement with the large $n$ expansion of the anomalous dimension
up to four-loop order which is available in \cite{Beccaria:2007pb}. However, since 
the $g^8$ contribution to \eqref{gamma large n} is quite lengthy, we merely 
give the first three loop orders. 

Interestingly, equation \eqref{eq:large n dense} can also be solved at large values
of the coupling by making use of the strong coupling expansion for twist operators in 
the $\mathfrak{sl}(2)$ sector \cite{BasKorKot08}. We defer the strong-coupling 
analysis to Appendix \ref{app:strong coupling} and only present the final result 
\beqa \nn
\gamma(L,n,m)&=&  \Big(4g-\frac{3\log 2}{\pi}\Big)\log\frac{n}{g}+6\,g\,(\log 2-1)+(1-L)+\frac{2}{m-1}+\frac{m}{2}\\
&&
+\frac{9\log 2}{\pi}-\frac{9(\log 2)^2}{2\pi}+\Op \Big(\frac{1}{g}\Big) \,.
\eeqa
%
\section{Outlook}

The appearance of the cusp anomalous dimension $f(g)$ and the virtual scaling 
function $B_L(g)$ beyond the $\mathfrak{sl}(2)$ sector in \eqref{mainresult} is quite 
remarkable and certainly promotes their universality. For further sub-leading corrections 
in spin, the endpoints of the effective condensate may not be enough to completely determine 
the corresponding contribution. In this case the remaining roots should also be taken into 
account and it is questionable if the scaling still resembles the behaviour of twist 
operators.

It will be quite interesting to investigate if the 
generalised scaling function $f(g,j)$ of the $\mathfrak{sl}(2)$ sector 
\cite{Freyhult:2007pz} also appears in the refined limit $S \to \infty$, $L \to \infty$ 
with $j=\sfrac{L}{\log S}$ fixed. 

In similarity to the known solvable cases of twist-two and three operators, 
it would be very interesting to see whether it is possible to construct higher loop 
contributions to the anomalous dimension \eqref{eq:one loop gamma} for general values 
of $m$ and $n$. A first step in this direction has been made in \cite{Beccaria:2007pb},
where the case of $m=2$ was analysed.

Furthermore, the decoupling procedure described in detail in \Appref{app:A} is based
on iteratively splitting one bosonic node of the corresponding Dynkin diagram into two 
fermionic ones. Although this is straightforward at the level of the equations, it remains 
obscure to us what the corresponding algebraic interpretation might be.
\subsection*{Acknowledgments}

We thank Matteo Beccaria for his collaboration at an early stage
of this project. We would especially like to thank  A.~Tseytlin for many discussions and  
A.~Tseytlin and A.~Tirziu for sharing their paper prior to publication with us. We have 
also benefited from
discussions with T.~Bargheer, L.~Dixon, K.~J.~Larsen, T.~{\L}ukowski, T.~McLoughlin, 
J.~Minahan, C.~Meneghelli and M.~Staudacher. This research was supported in part by the Swedish research 
council (VR). A.R. is supported by a STFC postdoctoral fellowship.
\newpage
\appendix
\section{The analytic one-loop solution}\label{app:A}
In this section we will study the following class of length-three operators
\begin{equation}\label{the operator}
\Tr \fldD^{n+m} \dot{\bar{\fldD}}^{m} \fldZ^3 \,.
\end{equation}
For $m=0$ this set reduces to twist-three operators of the $\mathfrak{sl}(2)$ 
sub-sector. Also the case of $m=1$ is redundant as \eqref{the operator} then
corresponds to descendents of twist-two operators. Therefore in 
what follows we will assume $m>1$. 
A subgroup of these operators for $m=2$ has been 
studied in \cite{Beccaria:2007pb} up to four-loop order. At one-loop order, on the 
other hand, a closed expression for the anomalous dimension of the ground states has 
been conjectured in \cite{BeiBiaMorSam04} for \textit{any} value of $m$ and $n$, 
see \eqref{eq:one loop gamma}. It is rather straightforward to prove this 
formula\footnote{M. Beccaria, private communication.} using the analytic solution 
provided below.

The excitation pattern for the higher-loop
Dynkin diagram in the upper part of \figref{Dynkindiagram} reads
\begin{equation}\label{pattern}
(K_1,K_2,K_3,K_4,K_5,K_6,K_7)=(0,0,n+2m-1,n+2m,n+2(m-1),m-1,0)\, ,
\end{equation}
where $K_\nu$ denotes the excitation number of the $\nu$-th node
of the Dynkin diagram. After a dualisation of the $u_3$ roots the corresponding one-loop 
system of equations describing this class of operators is given by
\begin{eqnarray}\label{first}
\left(\frac{u_{4,k}+\frac{i}{2}}{u_{4,k}-\frac{i}{2}}\right)^{3}&=&
\prod^{n+2m-2}_{j=1} 
 \frac{u_{4,k}-u_{5,j}-\frac{i}{2}}{u_{4,k}-u_{5,j}+\frac{i}{2}}\\
1&=&\prod^{m-1}_{j=1}\frac{u_{5,k}-u_{6,j}+\frac{i}{2}}{u_{5,k}-u_{6,j}-\frac{i}{2}} \prod^{n+2m}_{j=1}\frac{u_{5,k}-u_{4,j}-\frac{i}{2}}{u_{5,k}-u_{4,j}+\frac{i}{2}}\\ \label{last}
1&=&\prod^{m-1}_{\substack{j=1\\ j \neq k}} \frac{u_{6,k}-u_{6,j}+i}{u_{6,k}-u_{6,j}-i} 
\prod^{n+2m-2}_{j=1} \frac{u_{6,k}-u_{5,j}-\frac{i}{2}}{u_{6,k}-u_{5,j}+\frac{i}{2}}\, .
\end{eqnarray}
This set of Bethe equations is valid for $m\geq 1$. In the following section we will 
solve it \textit{exactly} thanks to a hidden recurrence relation 
between roots for different value of $m$.
%
\subsection{The one-loop recurrence}
We start by dualizing equation \eqref{first}. The system of equations 
\eqref{first}-\eqref{last} reduces to
\begin{eqnarray}
\left(\frac{u_{5,k}+i}{u_{5,k}-i}\right)^3&=&
\prod^{n+2m-2}_{\substack{j=1 \\ j\neq k}} \frac{u_{5,k}-u_{5,j}-i}{u_{5,k}-u_{5,j}+i}
\prod^{m-1}_{j=1} \frac{u_{5,k}-u_{6,j}+\frac{i}{2}}{u_{5,k}-u_{6,j}-\frac{i}{2}} 
\label{eq:rec1a}\\
1&=&\prod^{m-1}_{\substack{j=1 \\ j\neq k}} \frac{u_{6,k}-u_{6,j}-i}{u_{6,k}-u_{6,j}+i} 
\prod^{n+2m-2}_{j=1} \frac{u_{6,k}-u_{5,j}+\frac{i}{2}}{u_{6,k}-u_{5,j}-\frac{i}{2}} 
\label{eq:rec1b}\,.
\end{eqnarray}
The derivation of the hidden recurrence is based on the observation that the last equation 
is equivalent to a system of two coupled equations
\begin{eqnarray} 
1&=&\prod^{m-2}_{j=1}\frac{u_{6,k}-u_{7,j}-\frac{i}{2}}{u_{6,k}-u_{7,j}+\frac{i}{2}} 
\prod^{n+2m-2}_{j=1} \frac{u_{6,k}-u_{5,j}+\frac{i}{2}}{u_{6,k}-u_{5,j}-\frac{i}{2}}
\label{eq:split1}\\
1&=&\prod^{m-1}_{j=1} \frac{u_{7,k}-u_{6,j}+\frac{i}{2}}{u_{7,k}-u_{6,j}-\frac{i}{2}} 
\label{eq:split2}\,,
\end{eqnarray}
where we have introduced a new set of auxiliary roots $u_7$. We now introduce the 
Baxter function for the $u_6$ roots and their dual counterpart $\widetilde{u}_6$
\begin{eqnarray}
R(u) &\equiv& \prod^{m-2}_{j=1}(u-u_{7,j}-\tfrac{i}{2})
\prod^{n+2m-2}_{j=1} (u-u_{5,j}+\tfrac{i}{2})-
\prod^{m-2}_{j=1}(u-u_{7,j}+\tfrac{i}{2})
\prod^{n+2m-2}_{j=1} (u-u_{5,j}-\tfrac{i}{2})
\nonumber\\
&=&c_{6}\prod^{m-1}_{j=1}(u-u_{6,j}) \prod^{n+2m-4}_{j=1} (u-\widetilde{u}_{6,j})
\label{Q6}\,.
\end{eqnarray}
It is straightforward to derive the following two relations
\begin{equation} \label{Q6dual1}
\frac{R(u_{5,k}+\frac{i}{2})}{R(u_{5,k}-\frac{i}{2})}=\prod^{n+2m-2}_{\substack{j=1\\j\neq k}} \frac{u_{5,k}-u_{5,j}+i}{u_{5,k}-u_{5,j}-i}=\prod^{m-1}_{j=1} \frac{u_{5,k}-u_{6,j}+\frac{i}{2}}{u_{5,k}-u_{6,j}-\frac{i}{2}}\prod^{n+2m-4}_{j=1} \frac{u_{5,k}-\widetilde{u}_{6,j}+\frac{i}{2}}{u_{5,k}-\widetilde{u}_{6,j}-\frac{i}{2}}\,,
\end{equation}
\begin{equation}\label{Q6dual2}
\frac{R(u_{7,k}+\frac{i}{2})}{R(u_{7,k}-\frac{i}{2})}=\prod^{m-2}_{\substack{j=1\\j\neq k}} \frac{u_{7,k}-u_{7,j}+i}{u_{7,k}-u_{7,j}-i}=\prod^{m-1}_{j=1} \frac{u_{7,k}-u_{6,j}+\frac{i}{2}}{u_{5,k}-u_{7,j}-\frac{i}{2}}\prod^{n+2m-4}_{j=1} \frac{u_{7,k}-\widetilde{u}_{6,j}+\frac{i}{2}}{u_{7,k}-\widetilde{u}_{6,j}-\frac{i}{2}}\,.
\end{equation}
With the help of \eqref{Q6dual1} we now rewrite \eqref{eq:rec1a} as
\begin{equation}
\left(\frac{u_{5,k}+i}{u_{5,k}-i}\right)^3 =\prod^{n+2m-4}_{j=1} \frac{u_{5,k}-\widetilde{u}_{6,j}-\frac{i}{2}}{u_{5,k}-\widetilde{u}_{6,j}+\frac{i}{2}}\,,
\end{equation}
and once again dualize this set of equations by defining the polynomial $Q_5(u)$ as
\begin{equation}
Q_5(u)\equiv (u+i)^3 \, \prod^{n+2m-4}_{j=1}(u-\widetilde{u}_{6,j}+\tfrac{i}{2})-
               (u-i)^3 \, \prod^{n+2m-4}_{j=1}(u-\widetilde{u}_{6,j}-\tfrac{i}{2}) 
= c_{5}\prod^{n+2m-2}_{j=1}(u-u_{5,j}).\label{Q5}
\end{equation}
The function $Q_5(u)$ obeys the relation
\begin{equation}\label{Q5dual}
\frac{Q_5(\widetilde{u}_{6,k}+\frac{i}{2})}{Q_5(\widetilde{u}_{6,k}-\frac{i}{2})}=
\left(\frac{\widetilde{u}_{6,k}+\frac{3}{2}i}{\widetilde{u}_{6,k}-\frac{3}{2}i}\right)^3 
\,\prod^{n+2m-4}_{\substack{j=1 \\ j\neq k}}
 \frac{\widetilde{u}_{6,k}-\widetilde{u}_{6,j}+i}{\widetilde{u}_{6,k}-\widetilde{u}_{6,j}-i}=
\prod^{n+2m-2}_{j=1} 
 \frac{\widetilde{u}_{6,k}-u_{5,j}+\frac{i}{2}}{\widetilde{u}_{6,k}-u_{5,j}-\frac{i}{2}}\,.
\end{equation}
Since the $\widetilde{u}_6$ roots also solve \eqref{eq:split1}, we can decouple the $u_5$ 
roots. Likewise, we use \eqref{Q6dual2} to rewrite \eqref{eq:split2} in terms of 
$\widetilde{u}_6$. The resulting set of equations thus reads
\begin{eqnarray} 
\left(\frac{\widetilde{u}_{6,k}+\frac{3}{2} i}{\widetilde{u}_{6,k}-\frac{3}{2} i}\right)^3
&=&
\prod^{n+2m-4}_{\substack{j=1 \\ j\neq k}}
 \frac{\widetilde{u}_{6,k}-\widetilde{u}_{6,j}-i}{\widetilde{u}_{6,k}-\widetilde{u}_{6,j}+i}
\prod^{m-2}_{j=1}
 \frac{\widetilde{u}_{6,k}-u_{7,j}+\frac{i}{2}}{\widetilde{u}_{6,k}-u_{7,j}-\frac{i}{2}} 
\label{eq:rec2a}\\
1 &=&\prod^{m-2}_{\substack{j=1 \\ j\neq k}}\frac{u_{7,k}-u_{7,j}-i}{u_{7,k}-u_{7,j}+i} 
\prod^{n+2m-4}_{j=1} 
\frac{u_{7,k}-\widetilde{u}_{6,j}+\frac{i}{2}}{u_{7,k}-\widetilde{u}_{6,j}-\frac{i}{2}} 
\label{eq:rec2b}\,.
\end{eqnarray}
Comparing the two systems of equations, namely \eqref{eq:rec1a}-\eqref{eq:rec1b} with 
\eqref{eq:rec2a}-\eqref{eq:rec2b}, we note that the value of 
$m$ has been lowered by one, while the spin of the representation increased by 
$\tfrac{1}{2}$. Clearly, this procedure 
can be applied recursively until all second-level roots vanish and get {\it absorbed} 
into the spin representation of the first-level roots. Thus, after $(m-1)$ steps one 
can decouple \eqref{eq:rec1b} from \eqref{eq:rec1a} and one is left with a single system 
of equations\footnote{Please note, however, that in the penultimate step one should not 
perform the splitting, but rather dualize the nested set of equations directly.}. 
Before using this recursion to solve the system \eqref{eq:rec1a}-\eqref{eq:rec1b}, we 
will investigate the Baxter functions appearing in the intermediate steps.

Suppose that one has repeated the aforementioned procedure $\ell$ times. The intermediate 
equations then read

\begin{eqnarray}
\left(\frac{u^{(\ell)}_{k}+(1+\frac{\ell}{2}) i}{u^{(\ell)}_{k}-(1+\frac{\ell}{2}) i}\right)^3
&=&\prod^{n+2(m-1-\ell)}_{\substack{j=1 \\ j\neq k}} \frac{u^{(\ell)}_{k}-u^{(\ell)}_{j}-i}{u^{(\ell)}_{k}-u^{(\ell)}_{j}+i}\prod^{m-1-\ell}_{j=1} \frac{u^{(\ell)}_{k}-v^{(\ell)}_{j}+\frac{i}{2}}{u^{(\ell)}_{k}-v^{(\ell)}_{j}-\frac{i}{2}} 
\label{eq:recLa}\\
1 &=&\prod^{m-1-\ell}_{\substack{j=1 \\ j\neq k}} 
\frac{v^{(\ell)}_{k}-v^{(\ell)}_{j}-i}{v^{(\ell)}_{k}-v^{(\ell)}_{j}+i} 
\prod^{n+2(m-1-\ell)}_{j=1} 
\frac{v^{(\ell)}_{k}-u^{(\ell)}_{j}+\frac{i}{2}}{v^{(\ell)}_{k}-u^{(\ell)}_{j}-\frac{i}{2}}\,, 
\label{eq:recLb}
\end{eqnarray}
with the initial values
\begin{equation}
u^{(0)}_{k}\equiv u_{5,k}\,, \qquad u^{(1)}_{k}\equiv \widetilde{u}_{6,k}\,,
\end{equation}
\begin{equation}
v^{(0)}_{k}\equiv u_{6,k}\,, \qquad v^{(1)}_{k}\equiv u_{7,k}\,.
\end{equation}
A single step of the iteration relates the polynomials
\begin{equation}
P_{\ell}(u)\equiv \prod^{n+2(m-1-\ell)}_{j=1} (u-u^{(\ell)}_{j})\,,
\end{equation}
with consecutive values of $\ell$ through
\begin{equation} \label{recc}
P_{\ell}(u)=\left(u+(1+\tfrac{\ell}{2})\,i \right)^3 \,P_{\ell+1}(u+\tfrac{i}{2})-
            \left(u-(1+\tfrac{\ell}{2})\,i \right)^3 \,P_{\ell+1}(u-\tfrac{i}{2})\,.
\end{equation}
The initial polynomials \eqref{Q6} and \eqref{Q5} are respectively the second and the 
third member of this family, i.e.~$\widetilde{Q}_6(u)=P_1(u)$ and $Q_5(u)=P_0(u)$. 
It should be clear that the first element is given by
\begin{equation}
P_{-1}(u)=Q_{4}(u)=c_{4} \prod^{n+2m}_{j=1} (u-u_{4,j})\,.
\end{equation}

A general solution to the recurrence relation \eqref{recc} is given by
\begin{eqnarray}\label{reccSOL}
P_{\ell}(u)=\sum^{n}_{k=0} (-1)^k \binom{n}{k} 
\prod^{k}_{j=1} \left(u-\frac{(2j+\ell)i}{2}\right)^3
\prod^{n-k}_{j=1} \left(u+\frac{(2j+\ell)i}{2}\right)^3 
P_{\ell+n}\left(u+\left(\frac{n}{2}-k\right)i\right)\,,
\nonumber\\
\end{eqnarray}
with $n$ being an arbitrary positive integer. 

As already mentioned before, for $\ell=m-1$
equations \eqref{eq:recLa}-\eqref{eq:recLb} decouple and one is left
with a single equation
\begin{equation} \label{eq:decouple}
\left(\frac{u^{(m-1)}_{k}+\frac{m+1}{2} i}{u^{(m-1)}_{k}-\frac{m+1}{2} i}\right)^3
=\prod^{n}_{\substack{j=1 \\ j\neq k}} 
\frac{u^{(m-1)}_{k}-u^{(m-1)}_{j}-i}{u^{(m-1)}_{k}-u^{(m-1)}_{6}+i}\,. 
\end{equation}
It is noteworthy that these are the Bethe equations of a non-compact $\alg{sl}(2)$ 
magnet in the spin-$(-\frac{m+1}{2})$ representation. Thus, we have proven an equivalence 
between both systems noticed in \cite{BeiBiaMorSam04}. The 
corresponding Baxter equation can be solved exactly for the ground state. Please refer 
to \Appref{app:HighSpinSl2} for further details. The solution is the Wilson polynomial
\begin{eqnarray}
F_{n,m}(u) &\equiv& P_{m-1}(u)=c_{m-1} \prod^{n}_{j=1} (u-u^{(m-1)}_{j})
\nonumber\\
&=&{}_4 F_3\left(\left. \begin{array}{c}
-\frac{n}{2},\ \frac{n}{2}+1+\frac{3\,m}{2},\ \frac{1}{2}+iu,\ \frac{1}{2}-iu \\
1+\frac{m}{2},\  1+\frac{m}{2},\  1+\frac{m}{2}
\end{array}
\right| 1\right)\,.
\label{4F3}
\end{eqnarray}

Plugging this into \eqref{reccSOL}, one finds an explicit solution for the primary roots
\begin{equation}\label{Q4sol}
Q_4(u)=\sum^{m}_{k=0} (-1)^k \binom{m}{k} \prod^{k}_{j=1} 
\left(u-\tfrac{(2j-1)i}{2}\right)^3 \prod^{m-k}_{j=1} 
\left(u+\tfrac{(2j-1)i}{2}\right)^3 F_{n,m}
\left(u+\left(\tfrac{m}{2}-k\right)i\right)\,,
\end{equation}
while the $u_5$ roots are generated by
\begin{equation}\label{Q5sol}
Q_5(u)=\sum^{m-1}_{k=0} (-1)^k \binom{m-1}{k} 
\prod^{k}_{j=1} ( u-j\,i)^3 \prod^{m-1-k}_{j=1} (u+j\,i)^3 
\, F_{n,m}\left(u+\left(\tfrac{m-1}{2}-k\right)i\right)\,.
\end{equation}
\begin{figure}[t]\label{fig:rec}
\begin{minipage}{260pt}
\begin{picture}(260,260)(-80,-20)
\setlength{\unitlength}{1.2pt}
\small\thicklines
\multiput(65,  0)(40, 00){2}{\circle{15}}
\put     (73, 00){\line(1,0){24}}
\put( 65,10){\makebox(0,0)[b]{$u^{(p+1)}$}}
\put(105,10){\makebox(0,0)[b]{$v^{(p+1)}$}}
\multiput(130,20)(40, 00){3}{\circle{15}}
\multiput(165,25)(40, 00){2}{\line(1,-1){10}}
\multiput(165,15)(40, 00){2}{\line(1,1){10}}
\multiput(138,20)(40, 00){2}{\line(1,0){24}}
\put(130,30){\makebox(0,0)[b]{$u^{(p)}$}}
\put(170,30){\makebox(0,0)[b]{$v^{(p)}$}}
\put(210,30){\makebox(0,0)[b]{$v^{(p+1)}$}}
\multiput(65, 40)(40, 00){2}{\circle{15}}
\put     (73, 40){\line(1,0){24}}
\put(65,50){\makebox(0,0)[b]{$u^{(p)}$}}
\put(105,50){\makebox(0,0)[b]{$v^{(p)}$}}
\put      ( 85,56){\makebox(0,0)[b]{$\vdots$}}
\put      (140,56){\makebox(0,0)[b]{$\vdots$}}
%
%
\multiput( 65, 80)(40, 00){2}{\circle{15}}
\put     ( 73, 80){\line(1,0){24}} 
\put     ( 65, 90){\makebox(0,0)[b]{$\widetilde{u}_6$}}
\put     (105, 90){\makebox(0,0)[b]{$u_7$}}

\multiput(130,100)(40, 00){3}{\circle{15}}
\multiput(165,105)(40, 00){2}{\line(1,-1){10}}
\multiput(165, 95)(40, 00){2}{\line(1,1){10}}
\multiput(138,100)(40, 00){2}{\line(1,0){24}}
\put(130,110){\makebox(0,0)[b]{$u_5$}}
\put(170,110){\makebox(0,0)[b]{$u_6$}}
\put(210,110){\makebox(0,0)[b]{$u_7$}}
\multiput(65,120)(40, 00){2}{\circle{15}}
\put     (73,120){\line(1,0){24}} 
\put(65,130){\makebox(0,0)[b]{$u_5$}}
\put(105,130){\makebox(0,0)[b]{$u_6$}}
\multiput(65,160)(40,00){3}{\circle{15}}
\multiput(73,160)(40,00){2}{\line(1,0){24}}
\multiput(60,165)(40,00){2}{\line(1,-1){10}}
\multiput(60,155)(40,00){2}{\line(1,1){10}}
\put(65,170){\makebox(0,0)[b]{$u_4$}}
\put(105,170){\makebox(0,0)[b]{$u_5$}}
\put(145,170){\makebox(0,0)[b]{$u_6$}}
\dottedline{2}(111,  6)(123,14)
\dottedline{2}(111, 46)(123,54)
\dottedline{2}(111, 86)(123,94)
\dottedline{2}(123, 26)(111,34)
\dottedline{2}(123, 66)(111,74)
\dottedline{2}(123,106)(111,114)
\thinlines
\put(38,-03){\makebox(0,0)[b]{\small{$p+1$}}}
\qbezier[50](55,5)(40,20)(55,35)
\qbezier[50](95,5)(80,20)(95,35)
\put(38,37){\makebox(0,0)[b]{\small{$p$}}}
%
\put(38,77){\makebox(0,0)[b]{\small{$p=+1$}}}
%
\qbezier[50](55,85)(40,100)(55,115)
\qbezier[50](95,85)(80,100)(95,115)
\put(38,117){\makebox(0,0)[b]{\small{$p=0$}}}
%
\qbezier[50](55,125)(40,140)(55,155)
\put(38,157){\makebox(0,0)[b]{\small{$p=-1$}}}
\put(85,-20){\makebox(0,0)[b]{$\vdots$}}
\end{picture}
\end{minipage}
\caption{A schematic representation of the decoupling procedure. Dotted lines indicate the 
bosonic splitting and the dualization of the interjacent roots. The curved lines denote 
two kinds of recurrence relations $P$ and $K$, respectively. The exact one-loop solutions 
$Q_4$ and $Q_5$  belong to $P$ iteration, while $Q_6$ is part of the $K$ recurrence.}
\end{figure}
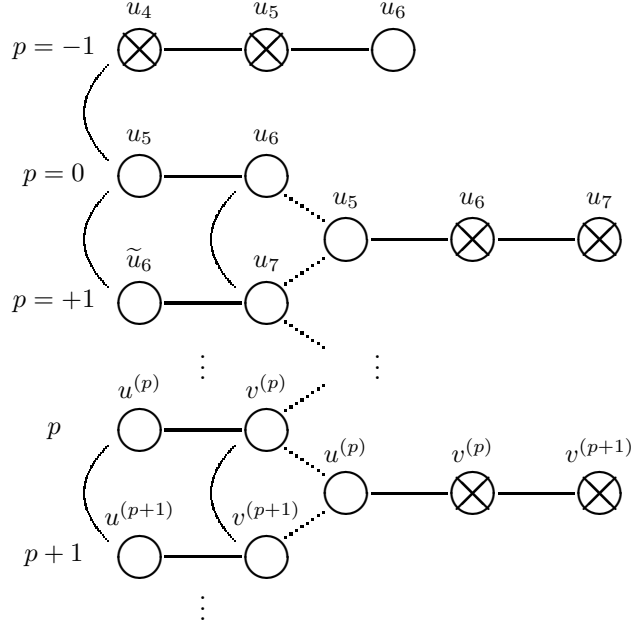
The Baxter function for the $u_6$ roots is a part of a \textit{different} recursive  
scheme. Defining
\begin{equation}
K_{\ell}(u) \equiv \prod^{m-1-\ell}_{j=1}(u-v^{(\ell)}_{j})\,,
\end{equation}
we find that $K_{\ell}$ obeys the following functional relation
\begin{equation}\label{eq:Krec}
K_{\ell}(u)=f_{\ell}(u)\,K_{\ell+1}\left(u+\tfrac{i}{2} \right)-
  \widetilde{f}_{\ell}(u)\,K_{\ell+1}\left(u-\tfrac{i}{2} \right)\,,
\end{equation}
with
\begin{equation}\label{effe}
f_\ell(u)=-\frac{P_\ell \left(u-\frac{i}{2}\right)}{P_{\ell+1}(u)} 
\qquad \mathrm{and} \qquad 
\widetilde{f}_\ell(u)=-\frac{P_\ell \left(u+\frac{i}{2}\right)}{P_{\ell+1}(u)}\,.
\end{equation}
A general solution to equation \eqref{eq:Krec} is presented in Appendix \ref{app:RecSol}. 
We fix the boundary conditions by specifying
\begin{equation}
K_{m-1}=1 \qquad \mathrm{and} \qquad K_{0}(u)\equiv Q_6(u)=\prod^{m-1}_{j=1}(u-u_{6,j})\,.
\end{equation}
Thus, we find that $Q_6$ is given by the following expression
\begin{eqnarray}
Q_{6}(u)&=&\prod^{m-2}_{k=0} f_{k}\left(u+\frac{k}{2}\,i\right)\nonumber\\
&&+\sum^{m-1}_{r=1}(-1)^r\, \sum^{m-2}_{j_1=0} \, \sum^{j_1-1}_{j_2=0} \, 
\ldots \sum^{j_{r-1}-1}_{j_r=0}\prod^{r}_{s=1} 
\widetilde{f}_{j_s}\left(u+\frac{j_s-2(r-s)}{2}\,i\right)\,
\prod^{j_r-1}_{k=0}f_{k}\left(u+\frac{k}{2}\,i\right)\nonumber\\
&&\times \prod^{r}_{s=2}\, \prod^{j_{s-1}-1}_{k=j_{s}+1}f_{k}
\left(u+\frac{k-2(r-s+1)}{2}\,i\right)\, 
\prod^{m-2}_{k=j_1+1}f_{k}\left(u+\frac{k-2r}{2}\,i\right)\,, 
 \label{Q6sol}
\end{eqnarray}
with $f_{p}(u)$ and $\widetilde{f}_{p}(u)$ given in \eqref{effe} and 
\begin{eqnarray}
P_{\ell}(u)&=&\sum^{m-1-\ell}_{k=0} (-1)^k \binom{m-1-\ell}{k}\, 
\prod^{k}_{j=1} \left(u-\frac{(2j+\ell)\,i}{2}\right)^3 \, 
\prod^{m-1-\ell-k}_{j=1} \left(u+\frac{(2j+\ell)\,i}{2}\right)^3 \nonumber\\
&& \times F_{n,m}\left(u+\left(\frac{m-1-\ell}{2}-k\right)i\right) \label{sol}\,,
\end{eqnarray}
as follows from \eqref{reccSOL} with $\ell+n=m-1$. 

\medskip

The complete solution of the one-loop problem is hence given by \eqref{Q4sol}, \eqref{Q5sol} and \eqref{Q6sol}. 
%

\subsection{Solution to the higher spin magnet}\label{app:HighSpinSl2}

The Bethe equations in \eqref{eq:decouple} correspond to an integrable non-compact 
$XXX_{-\frac{m+1}{2}}$ magnet. The Baxter equation associated with $XXX_{-s}$ magnets takes the following generic form
\begin{equation}\label{eq:sl2Baxter}
 (u+is)^L Q(u+i)+(u-is)^L Q(u-i)=t_L(u)Q(u)\,.
\end{equation}
{}For the special case of $L=3$, the transfer matrix of the ground states can be  
determined explicitly
\begin{equation}\label{lengththreetm}
 t_3(u)=2u^3+q_2u+q_3 \,, \quad q_2=-(n^2-n+6\,n\,s+6\,s^2)\,, \quad q_3=0\,.
\end{equation}
The solution to \eqref{eq:sl2Baxter} with \eqref{lengththreetm} can be found by noting 
that the Wilson polynomials $W_k$\,,
\begin{equation}
 \frac{W_k(u^2; a, b, c, d)}{(a+b)_k\,(a+c)_k\,(a+d)_k} \equiv 
       {}_4\,F_3\left(\left. \begin{array}{c}
            -k,\ k+a+b+c+d-1,\ a+i\,u,\ a-i\,u\\
            a+b,\ a+c,\ a+d
\end{array}
\right| 1\right)\,,
\end{equation}
satisfy the following difference equation (see e.g.~\cite{KoeSwa98})
\begin{equation}\label{eq:Wilson difference}
 k(k+a+b+c+d-1)y(u)=B(u)y(u+i)-[B(u)+D(u)]y(u)+D(u)y(u-i)\,.
\end{equation}
Here, $y(u)=W_k(u^2;a,b,c,d)$ and
\begin{equation}
 B(u)=\frac{(a-iu)(b-iu)(c-iu)(d-iu)}{2iu(2iu-1)}\,,\;
 D(u)=\frac{(a+iu)(b+iu)(c+iu)(d+iu)}{2iu(2iu+1)}\,.
\end{equation}
It is easy to check that \eqref{eq:sl2Baxter} is embedded in \eqref{eq:Wilson difference} 
upon the identification 
\begin{equation}
 a=\frac{1}{2} \,, \quad b=c=d=s \quad \mathrm{and} \quad k=\frac{n}{2}\,.
\end{equation}
Hence, the solution is given by
\begin{equation}
 Q(u)={}_4\,F_3\left(\left. 
    \begin{array}{c}
            -\tfrac{n}{2},\ \tfrac{n}{2}+3s-\tfrac{1}{2},\ \tfrac{1}{2}+i\,u,\ \tfrac{1}{2}-i\,u\\
            \tfrac{1}{2}+s,\ \tfrac{1}{2}+s,\ \tfrac{1}{2}+s
    \end{array}
\right| 1\right)\,.
\end{equation}
For the special value of $s=\tfrac{m+1}{2}$ one finds \eqref{4F3}.
\subsection{Solution of the recurrence}\label{app:RecSol}
The class of functional equations
\begin{equation}
A_p (u)=f_p (u)\,A_{p+1}\left(u+\frac{i}{2} \right)
       -\widetilde{f}_p (u)\,A_{p+1}\left(u-\frac{i}{2} \right)
\end{equation}
is solved by 
\begin{eqnarray} \nonumber
A_{p}(u)&=&\prod^{n-1}_{k=0} f_{p+k}\left(u+\frac{k}{2}\,i\right)
                             A_{p+n}\left(u+\frac{n}{2}\,i \right)\nonumber\\
&&+\sum^{n}_{r=1}(-1)^r\, \sum^{n-1}_{j_1=0} \, \sum^{j_1-1}_{j_2=0} \, \ldots \sum^{j_{r-1}-1}_{j_r=0}\prod^{r}_{s=1} \widetilde{f}_{p+j_s}\left(u+\frac{j_s-2(r-s)}{2}\,i\right)\,
\prod^{j_r-1}_{k=0}f_{p+k}\left(u+\frac{k}{2}\,i\right) \nonumber\\
&&\times \prod^{r}_{s=2}\, \prod^{j_{s-1}-1}_{k=j_{s}+1}f_{p+k}\left(u+\frac{k-2(r-s+1)}{2}\,i\right)\, \prod^{n-1}_{k=j_1+1}f_{p+k}\left(u+\frac{k-2r}{2}\,i\right) \nonumber\\
&&\times A_{p+n}\left(u+\frac{n-2\,r}{2}\,i \right) \label{rloesung}\,.
\end{eqnarray}
The proof is by induction. 
\section{The strong coupling limit}\label{app:strong coupling}

In order to analyse the sub-leading contribution to \eqref{eq:large n gamma} we 
decompose the density $\hat{\sigma}(t)$ into its parity even and odd parts
\begin{equation}
 \hat{\sigma}(t)\frac{e^t-1}{t}=\frac{\gamma_+(2gt)}{2gt}+\frac{\gamma_-(2gt)}{2gt} \,.
\end{equation}
Since the kernels $K_{0}$ and $K_1$ in \eqref{eq:kernels} are given by the sum over 
Bessel functions, the functions $\gamma_{\pm}$ take the form of a Neumann series
\begin{equation}
\gamma_+(2gt)=2\sum_{n=1}^\infty 2nJ_{2n}(2gt)\gamma_{2n}\,,
\quad
\gamma_-(2gt)=2\sum_{n=1}^\infty (2n-1)J_{2n-1}(2gt)\gamma_{2n-1} \,.
\end{equation}
Using this decomposition of the fluctuation density it is possible to rewrite the integral 
equation \eqref{eq:large n dense} as an infinite system of equations with $n\ge 1$
\begin{eqnarray}\label{largendecomposed}
&&\int_0^\infty \frac{dt}{t}\left(\frac{\gamma_+(t)}{1-e^{-t/2g}}
-\frac{\gamma_-(t)}{e^{t/2g}-1}\right)J_{2n}(t)=
\frac{L}{8ng}+h_{2n}-
\frac{1}{2}\int_0^\infty dt\frac{J_{2n}(2gt)}{2gt}e^{-t\frac{m-1}{2}} \,, \nonumber\\
&&\int_0^\infty \frac{dt}{t}\left(\frac{\gamma_-(t)}{1-e^{-t/2g}}+
\frac{\gamma_+(t)}{e^{t/2g}-1}\right)J_{2n-1}(t)=h_{2n-1} \,,
\end{eqnarray}
where the term $h_n=h_n(g)$ is given by the expression
\begin{equation}
h_n=\frac{1}{4}\int_0^\infty dt\left(\frac{3}{e^t-1}-
\frac{e^{-(m+1)t/2}}{1-e^{-t}}-
\frac{L-3}{e^{t/2}+1}\right)\left(\frac{J_n(2gt)}{gt}-\delta_{n,1}\right) \,.
\end{equation}
Since the left hand side of the equations \eqref{largendecomposed} is the same as in 
the case of the BES equation, 
we expect to be able to express the solution in terms of the solution to 
the BES equation. For this purpose we introduce a new parameter $j$, which interpolates 
between the system corresponding to the BES equation  and \eqref{largendecomposed}
\begin{eqnarray}
&&\int_0^\infty \frac{dt}{t}\left(\frac{\gamma_+(t,j)}{1-e^{-t/2g}}-
\frac{\gamma_-(t,j)}{e^{t/2g}-1}\right)J_{2n}(t)=
\frac{j L}{8ng}+jh_{2n}-\frac{j}{2}\int_0^\infty dt\frac{J_{2n}(2gt)}{2gt}
e^{-t\frac{m-1}{2}} \,, \nonumber\\
&&\int_0^\infty \frac{dt}{t}\left(\frac{\gamma_-(t,j)}{1-e^{-t/2g}}
+\frac{\gamma_+(t,j)}{e^{t/2g}-1}\right)J_{2n-1}(t)
=jh_{2n-1}+\frac{1}{2}(1-j)\delta_{n,1} \,.
\end{eqnarray}
Setting $j=0$ gives back the BES equation while $j=1$ corresponds to 
\eqref{largendecomposed}. 
Multiplying the first equation by 
$(2n)\gamma_{2n}(t,j')$ and the second by $(2n-1)\gamma_{2n-1}(t,j')$, 
summing over all $n$ and finally subtracting the two equations leads to a left 
hand side that is symmetric under exchange of $j$ and $j'$, see \cite{BasKor09} 
for details. Using this fact and setting $j=0$ and $j'=1$, we find
\begin{eqnarray}
\nonumber\gamma_1(g,1)&=&
\frac{1}{4}\int_0^\infty dt\left(3-e^{-t(m-1)/4g}\right)
\left(\frac{\gamma_-(t,0)}{(e^{t/2g}-1)gt}+
\frac{\gamma_+(t,0)}{(e^{-t/2g}-1)gt}-\frac{\gamma_1(g,0)}{(e^{t/2g}-1)g}\right)\\
&&-\frac{L-3}{4}\int_0^\infty dt\left(\frac{\gamma_-(t,0)}{(e^{t/4g}+1)gt}+
\frac{\gamma_+(t,0)}{(e^{-t/4g}+1)gt}-\frac{\gamma_1(g,0)}{(e^{t/4g}+1)g}\right).
\end{eqnarray}
The finite order correction is then given by $16\,g^2\,\gamma_1(g,1)$. A change of 
variables as in \cite{BasKorKot08},
\begin{equation}
2\gamma_\pm(t,0)=\left(1-\mbox{sech}(\tfrac{t}{2g})\right)\Gamma_\pm(t,0)
\pm\tanh(\tfrac{t}{2g})\Gamma_\mp(t,0) \,,
\end{equation}
leads to
\begin{eqnarray}\label{gamma1(g,1)}
\nonumber \gamma_1(g,1)&=&\frac{1}{16g^2}(L-3)\epsilon_1(g)+\gamma_1(g,0)(L-3)\log 2+\frac{3}{2}B_2(g)\\
&&+\frac{1}{2}\int_0^\infty dt e^{-t(m+1)/4g}
\left(\frac{1}{4gt}\left(\Gamma_+(t,0)+\Gamma_-(t,0)\right)+
\frac{\gamma_1(g,0)}{(e^{t/2g}-1)}\right)\,.
\end{eqnarray}
At this stage, we make use of the solution of the BES equation obtained in 
\cite{BasKorKot08},
\begin{equation}
\nonumber\Gamma_+(t,0)=\sum_{k=0}^\infty (-1)^{k+1}J_{2k}(t)\Gamma_{2k} \,,
\quad
\Gamma_-(t,0)=\sum_{k=0}^\infty (-1)^{k+1}J_{2k-1}(t)\Gamma_{2k-1} \,,
\end{equation}
where the coefficients $\Gamma_k$ are given by
\begin{eqnarray}
\Gamma_k &=& -\frac{1}{2}\Gamma_k^{(0)}+
\sum_{p=1}^\infty\frac{1}{g^p}\left(c_p^-\Gamma_k^{(2p-1)}+c_p^+\Gamma_k^{2p}\right)\,,\\
\Gamma_{2m}^{(p)}&=&\frac{\Gamma(m+p-\tfrac{1}{2})}{\Gamma(m+1)\Gamma(\tfrac{1}{2})}\,, \quad
\Gamma_{2m-1}^{(p)}=\frac{(-1)^p\Gamma(m-\tfrac{1}{2})}{\Gamma(m+1-p)\Gamma(\tfrac{1}{2})} \,.
\end{eqnarray}
The prefactors $c_p^{\pm}$ explicitly depend on $g$ and can be
determined from the so-called all-loop quantization condition of \cite{BasKorKot08}. 
Bearing that in mind, we find for the integral in \eqref{gamma1(g,1)} 
\begin{eqnarray}
I(g)&=&\nonumber\frac{1}{2}\int_0^\infty dt \, e^{-t(m+1)/4g}
\left(\frac{1}{4gt}\left(\Gamma_+(t,0)+\Gamma_-(t,0)\right)+
\frac{\gamma_1(g,0)}{(e^{t/2g}-1)}\right)\\
\nonumber&=&\frac{1}{8g}\sum_{k=1}^\infty(-1)^{k+1}
\left(\left(\frac{ig}{x(i\tfrac{m-1}{2})}\right)^{2k}
\frac{\Gamma_{2k}}{2k}+\left(\frac{ig}{x(i\tfrac{m-1}{2})}\right)^{2k-1}
\frac{\Gamma_{2k-1}}{2k-1}\right)\\
&&+\frac{\gamma_1(g,0)}{4g}\int_0^\infty dt \, \frac{e^{-t(m-1)/4g}}{e^{t/2g}-1}
-\frac{1}{8g}\int_0^\infty dt \, e^{-t(m-1)/4g}\frac{J_0(t)}{t} \, \Gamma_0 \nonumber\\
&&+\frac{1}{8g}\int_0^\infty dt \, e^{-t(m-1)/4g}\frac{J_{1}(t)}{t} \, \Gamma_{-1} \,.
\end{eqnarray}
Subsequently, using that $\Gamma_0=4\,g\,\gamma_1(g,0)$ and $\Gamma_{-1}=1$, we can 
recast $I(g)$ as 
\begin{eqnarray}\label{newintegral}
\nonumber I(g) &=& \frac{1}{8g}\sum_{k=1}^\infty(-1)^{k+1}
\left(\left(\frac{ig}{x(i\tfrac{m-1}{2})}\right)^{2k}
\frac{\Gamma_{2k}}{2k}+\left(\frac{ig}{x(i\tfrac{m-1}{2})}\right)^{2k-1}
\frac{\Gamma_{2k-1}}{2k-1}\right)\\
&+&\frac{\gamma_1(g,0)}{4g}\int_0^\infty dt \, e^{-t(m-1)/4g}
\left(\frac{1}{e^{t/2g}-1}-2g\frac{J_0(t)}{t}\right)+\frac{1}{8g}
\frac{ig}{x(i\tfrac{m-1}{2})}\,.
\end{eqnarray}
Both sums and the integral in the above formula may be performed analytically leading to
\begin{equation}
I(g) = \frac{1}{8g^2(m-1)}+
\frac{f(g)}{32g^2}\left(\log g-\psi_0\left(\tfrac{m+1}{2}\right)+\tfrac{m-1}{4g}\right)\,.
\end{equation}
Hence we obtain
\begin{eqnarray}
\gamma_1(g,1)&=&\frac{1}{16g^2}(L-3)\epsilon_1(g)+\frac{f(g)}{16g^2}(L-3)\log 2+
\frac{3}{2}\frac{B_2(g)}{16g^2}+\frac{1}{8g^2(m-1)}\\
&+&\frac{f(g)}{32g^2}\left(\log g-\psi_0\left(\tfrac{m+1}{2}\right)+\tfrac{m-1}{4g}\right)
+\dots \,. \nonumber
\end{eqnarray}
The anomalous dimension for $m\geq2$ is consequently given by
\begin{eqnarray}
\nonumber \gamma(L,n)&=& f(g)\left(\log n+\tfrac{3}{2}\EulerGamma+
\tfrac{1}{2}\psi_0\left(\tfrac{m+1}{2}\right)-(L-3)\log2\right)+16g^2\gamma_1(g,1)\\
&=& \Big(4g-\frac{3\log 2}{\pi}\Big)\log\frac{n}{g}+6g(-1+\log 2)+(1-L)+\frac{2}{m-1}+\frac{m}{2}\nonumber\\
&&+\frac{9\log 2}{\pi}-\frac{9(\log 2)^2}{2\pi}+\Op \big(\tfrac{1}{g}\big) \,.
\end{eqnarray}


\begin{thebibliography}{99}
\bibitem{BelGorKor06}
  A.~V.~Belitsky, A.~S.~Gorsky and G.~P.~Korchemsky,
  Nucl.\ Phys.\  B {\bf 748}, 24 (2006)
  [arXiv:hep-th/0601112].

\bibitem{Collins:1989bt}
  J.~C.~Collins,
  Adv.\ Ser.\ Direct.\ High Energy Phys.\  {\bf 5} (1989) 573
  [arXiv:hep-ph/0312336].

\bibitem{BelKorPas09}
  A.~V.~Belitsky, G.~P.~Korchemsky and R.~S.~Pasechnik,
  Nucl.\ Phys.\  B {\bf 809}, 244 (2009)
  [arXiv:0806.3657 [hep-ph]].

\bibitem{KotRejZie09}
  A.~V.~Kotikov, A.~Rej and S.~Zieme,
  Nucl.\ Phys.\  B {\bf 813}, 460 (2009)
  [arXiv:0810.0691 [hep-th]].

\bibitem{BecBelKotZie09}
  M.~Beccaria, A.~V.~Belitsky, A.~V.~Kotikov and S.~Zieme,
  [arXiv:0908.0520 [hep-th]].

\bibitem{KotLip03}
 A.~V.~Kotikov and L.~N.~Lipatov,
 Nucl.\ Phys.\  B {\bf 661}, 19 (2003),
 Erratum-ibid.\  B {\bf 685}, 405 (2004),
 [arXiv:hep-ph/0208220].

\bibitem{MocVerVog04}
 S.~Moch, J.~A.~M.~Vermaseren and A.~Vogt,
 Nucl.\ Phys.\  B {\bf 688}, 101 (2004),
 [arXiv:hep-ph/0403192].

\bibitem{KotLipRejStaVel07}
  A.~V.~Kotikov, L.~N.~Lipatov, A.~Rej, M.~Staudacher and V.~N.~Velizhanin,
  J.\ Stat.\ Mech.\  {\bf 0710}, P10003 (2007)
  [arXiv:0704.3586 [hep-th]].

\bibitem{BajJanLuk08}
  Z.~Bajnok, R.~A.~Janik and T.~Lukowski,
  Nucl.\ Phys.\  B {\bf 816}, 376 (2009)
  [arXiv:0811.4448 [hep-th]].

\bibitem{FiaSanSieZan07}
  F.~Fiamberti, A.~Santambrogio, C.~Sieg and D.~Zanon,
  Phys.\ Lett.\  B {\bf 666}, 100 (2008)
  [arXiv:0712.3522 [hep-th]].
$\bullet$
  V.~N.~Velizhanin,
  [arXiv:0808.3832 [hep-th]].

\bibitem{BecForLukZie09}
  M.~Beccaria, V.~Forini, T.~Lukowski and S.~Zieme,
  JHEP {\bf 0903}, 129 (2009)
  [arXiv:0901.4864 [hep-th]].

\bibitem{FiaSanSie09}
  F.~Fiamberti, A.~Santambrogio and C.~Sieg,
  [arXiv:0908.0234 [hep-th]].

\bibitem{BeiEdeSta07}
  N.~Beisert, B.~Eden and M.~Staudacher,
  J.\ Stat.\ Mech.\  {\bf 0701}, P021 (2007)
  [arXiv:hep-th/0610251].

\bibitem{Bern:2006ew}
  Z.~Bern, M.~Czakon, L.~J.~Dixon, D.~A.~Kosower and V.~A.~Smirnov,
  Phys.\ Rev.\  D {\bf 75}, 085010 (2007)
  [arXiv:hep-th/0610248].

\bibitem{BasKorKot08}
  B.~Basso, G.~P.~Korchemsky and J.~Kotanski,
  Phys.\ Rev.\ Lett.\  {\bf 100} (2008) 091601
  [arXiv:0708.3933 [hep-th]].

\bibitem{KosSerVol08}
  I.~Kostov, D.~Serban and D.~Volin,
  JHEP {\bf 0808}, 101 (2008)
  [arXiv:0801.2542 [hep-th]].
 
\bibitem{Gubser:2002tv}
  S.~S.~Gubser, I.~R.~Klebanov and A.~M.~Polyakov,
  Nucl.\ Phys.\  B {\bf 636}, 99 (2002)
  [arXiv:hep-th/0204051].
\bibitem{Frolov:2002av}
  S.~Frolov and A.~A.~Tseytlin,
  JHEP {\bf 0206}, 007 (2002)
  [arXiv:hep-th/0204226].

\bibitem{Roiban:2007ju}
  R.~Roiban and A.~A.~Tseytlin,
  Phys.\ Rev.\  D {\bf 77}, 066006 (2008)
  [arXiv:0712.2479 [hep-th]].
$\bullet$
  R.~Roiban and A.~A.~Tseytlin,
  JHEP {\bf 0711}, 016 (2007)
  [arXiv:0709.0681 [hep-th]].

\bibitem{Korchemsky:1985xj}
  G.~P.~Korchemsky and A.~V.~Radyushkin,
  Phys.\ Lett.\  B {\bf 171}, 459 (1986).
  $\bullet$
  G.~P.~Korchemsky,
  Mod.\ Phys.\ Lett.\  A {\bf 4}, 1257 (1989).

\bibitem{Alday:2007hr}
  L.~F.~Alday and J.~M.~Maldacena,
  JHEP {\bf 0706}, 064 (2007)
  [arXiv:0705.0303 [hep-th]].

\bibitem{FreZie09}
  L.~Freyhult and S.~Zieme,
  Phys.\ Rev.\  D {\bf 79}, 105009 (2009)
  [arXiv:0901.2749 [hep-th]].

\bibitem{Freyhult:2007pz}
  L.~Freyhult, A.~Rej and M.~Staudacher,
  J.\ Stat.\ Mech.\  {\bf 0807}, P07015 (2008)
  [arXiv:0712.2743 [hep-th]].

\bibitem{FioGriRos08}
  D.~Bombardelli, D.~Fioravanti and M.~Rossi,
  Nucl.\ Phys.\  B {\bf 810}, 460 (2009)
  [arXiv:0802.0027 [hep-th]].

\bibitem{FioGriRos09}
  D.~Fioravanti, P.~Grinza and M.~Rossi,
  Phys.\ Lett.\  B {\bf 675}, 137 (2009)
  [arXiv:0901.3161 [hep-th]].

\bibitem{BecForTirTse08}
  M.~Beccaria, V.~Forini, A.~Tirziu and A.~A.~Tseytlin,
  Nucl.\ Phys.\  B {\bf 812}, 144 (2009)
  [arXiv:0809.5234 [hep-th]].

\bibitem{DixMagSte08}
  L.~J.~Dixon, L.~Magnea and G.~Sterman,
  JHEP {\bf 0808}, 022 (2008)
  [arXiv:0805.3515 [hep-ph]].

\bibitem{Kruczenski:2004wg}
  M.~Kruczenski,
  JHEP {\bf 0508}, 014 (2005)
  [arXiv:hep-th/0410226].

\bibitem{Dorey:2008vp}
  N.~Dorey and M.~Losi,
  [arXiv:0812.1704 [hep-th]].

\bibitem{Alday:2007mf}
  L.~F.~Alday and J.~M.~Maldacena,
  JHEP {\bf 0711}, 019 (2007)
  [arXiv:0708.0672 [hep-th]].

\bibitem{Kruczenski:2008bs}
  M.~Kruczenski and A.~A.~Tseytlin,
  Phys.\ Rev.\  D {\bf 77}, 126005 (2008)
  [arXiv:0802.2039 [hep-th]].

\bibitem{FreKruTir09}
  L.~Freyhult, M.~Kruczenski and A.~Tirziu,
  JHEP {\bf 0907}, 038 (2009)
  [arXiv:0905.3536 [hep-th]].

\bibitem{TirTse09}
  A.~Tirziu and A.~A.~Tseytlin,
 [arXiv:0911.2417 [hep-th]].

\bibitem{Beisert:2005fw}
  N.~Beisert and M.~Staudacher,
  Nucl.\ Phys.\  B {\bf 727}, 1 (2005)
  [arXiv:hep-th/0504190].

\bibitem{BeiBiaMorSam04}
  N.~Beisert, M.~Bianchi, J.~F.~Morales and H.~Samtleben,
  JHEP {\bf 0407}, 058 (2004)
  [arXiv:hep-th/0405057].

\bibitem{Eden:2006rx}
  B.~Eden and M.~Staudacher,
  J.\ Stat.\ Mech.\  {\bf 0611} (2006) P014,
  [arXiv:hep-th/0603157].

\bibitem{Korchemsky:1995be}
  G.~P.~Korchemsky,
  Nucl.\ Phys.\  B {\bf 462}, 333 (1996)
  [arXiv:hep-th/9508025].

\bibitem{BasKor09}
  B.~Basso and G.~P.~Korchemsky,
  Nucl.\ Phys.\  B {\bf 807}, 397 (2009)
  [arXiv:0805.4194 [hep-th]].

\bibitem{Fioravanti:2008rv}
  D.~Fioravanti, P.~Grinza and M.~Rossi,
  Nucl.\ Phys.\  B {\bf 810}, 563 (2009)
  [arXiv:0804.2893 [hep-th]].

\bibitem{Beccaria:2007pb}
  M.~Beccaria,
  JHEP {\bf 0709}, 023 (2007),
  [arXiv:0707.1574 [hep-th]].
$\bullet$
  M.~Beccaria and V.~Forini,
  JHEP {\bf 0806}, 077 (2008)
  [arXiv:0803.3768 [hep-th]].

\bibitem{Beccaria:2007vh}
  M.~Beccaria,
  JHEP {\bf 0706}, 054 (2007),
  [arXiv:0705.0663 [hep-th]].

\bibitem{KoeSwa98}
  R.~Koekoek, and R.~F.~Swarttouw, 
  {\it`` The Askey-Scheme of Hypergeometric Orthogonal Polynomials and its q-Analogue'',} 
  Delft, Netherlands: Technische Universiteit Delft, 
  Faculty of Technical Mathematics and Informatics Report {\bf 98-17}, p.24-26 1998.

\end{thebibliography}
\end{document}